\documentstyle[emulateapj]{article}

\slugcomment{Submitted to the Astrophysical Journal}

\lefthead{Wise \& Sarazin}
\righthead{Cold Gas in Cluster Cooling Flows}

\begin{document}

\title{\bf X--ray Absorption Due to Cold Gas in Cluster Cooling Cores}

\author{Michael W. Wise}
\affil{Massachusetts Institute of Technology, Center for Space Research \\
       Building NE80--6015, Cambridge, MA 02139--4307 \\
       E--mail: wise@space.mit.edu}

\and

\author{Craig L. Sarazin}
\affil{Department of Astronomy, University of Virginia \\
       P. O. Box 3818, Charlottesville, VA 22903--0818 \\
       E-mail: cls7i@virginia.edu}

\begin{abstract}
We have calculated the emergent X--ray properties for models of
cluster cooling flows including the effects of accumulated cooled
material.
The opacity of this cooled gas can reduce the overall X--ray
luminosity of the cooling flow, and values of $\dot M$ based on these
luminosities can underestimate the true value by factors of $\sim 2$.
We find that accumulated cooled material can produce emergent surface
brightness profiles much like those observed even for nearly
homogeneous gas distributions.
Consequently, much more of the gas may be cooling below
X--ray emitting temperatures in the central regions of cooling flows
($r \la 10$ kpc) than one would infer from observed X--ray surface
brightness profiles assuming the gas was optically thin. 
Similarly, the central densities and pressures in cooling flows may
have been underestimated.
We show that distributed absorption in cooling flows produces a
number of observable effects in the spectrum which may allow it to be
differentiated from absorption due to gas in our Galaxy.
These include a characteristic suppression of the continuum below
$\sim 2$ keV, absorption features such as a redshifted O K--edge, and
diminished intensity of resonance emission lines.
Spectra including the effects of intrinsic absorption are not well fit
by foreground absorbing models. 
Attempting to fit such models to the spatially resolved spectra can
lead to underestimates of the true absorbing column by factors of 3--20.   
Fits to integrated spectra of the entire cooling flow region can either
underestimate or overestimate the mass of the absorbing gas depending
on the specifics of the model.  
We discuss the potential detection of these effects with {\it AXAF},
{\it XMM}, and Astro-E. 
\end{abstract}


\section{Introduction}
\label{sec:intro}

More than half of the clusters of galaxies
observed at X--ray wavelengths
exhibit evidence for cool gas in their cores
(\markcite{edge92} Edge, Stewart, \& Fabian 1992).
The cooling time of this gas is often short compared to 
cluster ages; therefore, the gas cools at rates that are
often very large, $\dot M_c \sim 10-2000 ~M_\odot$ yr$^{-1}$.
In the absence of a heating mechanism to balance cooling, gravity
and pressure from the hotter, surrounding gas will drive the cooling
material into the core of the cluster.
If cooling flows are long--lived phenomena, these rates imply that 
$\sim 10^{12} ~$$M_{\sun}$ of material would cool over the lifetime of
the cluster.
Determining the final state of this cooling material remains
the fundamental problem concerning the cooling flow theory.

The most obvious repositories for the cooling gas are cold
molecular and atomic clouds, and stars. 
The central dominant galaxy in cooling flow clusters often have
blue stellar colors in their central regions, which indicate that
stars are currently being formed there (McNamara \& O'Connell 1992).
However, the observed star formation rates are generally $\la 10\%$ of
$\dot M_c$. Therefore star formation cannot account for the cooling
material without appealing to unusual initial mass functions. 
Similarly, extensive efforts have been made to detect the accumulated
cooled material either as gas at some temperature below the X--ray
emitting regime ($T \la 10^6$ K).
Gas is seen in cooling flows at $10^4$ K (Heckman et al.\ 1989; Baum 1992)
and in a few cases, as neutral atomic or molecular gas 
(Lazareff et al.\ 1989;  Mirabel, Sanders, \& Kazes 1989;
McNamara, Bregman, \& O'Connell 1990; Jaffe 1992;
O'Dea, Baum, \& Gallimore 1994a).
Dust is also seen in a number of cooling cores (Bregman, McNamara, \&
O'Connell 1990; Wise et al.\ 1993).
In all cases, however, the detected levels of atomic and molecular gas
are too low ($\la 10^9$--$10^{10} M_\odot$) to account for the cooled
gas which would accumulate over the age of the cluster.

The detection by White et al.\ (1991) of excess X--ray absorption
in a sample of cluster cooling flows was the first direct
evidence for a sufficiently large mass of cold material.
Using {\it Einstein} SSS (Solid State Spectrometer) spectra, these
authors found that many cooling flow clusters exhibit significant
levels of excess absorption over that expected from the Galaxy
with typical excess column densities of $10^{21}$ cm$^{-2}$.
Evidence for excess absorption in cooling flows has also been found in
analyses of X-ray spectra taken with
detectors on {\it Einstein}, {\it ROSAT}, {\it BBXRT}, and {\it ASCA}
(Lea, Mushotzky, \& Holt 1982;
Miyaji 1991;
Allen et al.\ 1993;
Fabian et al.\ 1994;
Allen \& Fabian 1997).
The excess absorption columns detected by White et al.\ (1991) 
were only observed in clusters with spectral evidence for cooling
flows and show some evidence for a correlation between $\dot M$ and
$\Delta N_H$. 
This correlation and the observed spatial coincidence between excess
absorption and cooling flows suggests that the absorbing material is
intrinsic to the cluster and probably located within the cooling
flow region. 
Integrating these excess column densities over the area of the cooling
region in cluster cores implies the presence of large quantities of
cold absorbing material (M$_{cold} \sim 10^{11}$--$10^{12} \dot M$)
and may represent the first direct evidence for the large amounts of
cooled material which current cooling flow models predict
(White et al.\ 1991).

On the other hand, very extensive searches have been made to detect
the excess absorber in emission or absorption at radio wavelengths in
lines of H~I or CO and have not succeeded (e.g., McNamara \& Jaffe
1993; Antonucci \& Barvainis 1994; O'Dea et al.\ 1994b).
It is becoming difficult to understand how so much X--ray absorbing
gas could have escaped detection in other wavebands 
(e.g., Voit \& Donahue 1995).

Often  the {\it ROSAT} PSPC spectra of cooling flows are inconsistent
with large columns of excess foreground absorption
(e.g., Sarazin, Wise, \& Markevitch 1998), but are consistent with and
may require large amounts of internal absorption in the cooling flow
(Allen \& Fabian 1997).
For nearby clusters where the {\it ROSAT} or {\it ASCA} observations
can resolve the cooling flow region, the excess absorption appears to
be concentrated to the center of the cluster and cooling flow 
($r \la 200$ kpc) (Allen et al.\ 1993; Irwin \& Sarazin 1995;
Fabian et al.\ 1994; Allen \& Fabian 1997).

In the standard data analysis of X--ray spectra (e.g., in {\sc XSPEC}
or {\sc IRAF/PROS}), it is conventional to treat absorbers as lying in the 
foreground of emission components. 
This assumption allows the simple arithmetic combination of additive 
emission sources and multiplicative absorption components.
However, X--ray observations suggest that the excess absorbing 
material in cluster cooling cores occupies the same spatial region 
as the X--ray emitting gas (Allen et al.\ 1993; Irwin \& Sarazin 1995;
Allen \& Fabian 1997).
Certainly, one would expect such a spatial correlation if the absorber
originated as cooling X--ray gas.
Including the effects of absorbing material which is intermixed with 
the X--ray emitting gas is not easily treated within the framework of 
conventional X--ray spectral modeling.
Allen \& Fabian (1997) used de-projection techniques based on {\it ROSAT}
PSPC X-ray colors to attempt to determine the three dimensional distribution
of the absorption.

In order to provide more accurate models of the X--ray spectra of
cooling flows and to assess the effect of intermixed absorbing material,
we have calculated the emergent X--ray properties for a set of
inhomogeneous cooling flow models including the opacity due to 
accumulated cooled gas.
For a given cooling flow model, we have self-consistently included
the X--ray absorption due to varying fractions of the total cooled 
material generated by the model.
The details of the input models and the numerical solution of the
radiative transfer equation are described in \S\ref{sec:calc}.
In \S\ref{sec:results}, we present X--ray luminosities, spectra,
integrated surface brightness profiles, and emission line equivalent widths,
and spectral profiles for the models.
The implications of these results for the interpretation of existing 
X--ray observations and their application to observations with
{\it AXAF} and Astro-E are discussed in \S\ref{sec:derive}.
In the final section (\S\ref{sec:conclude}), we summarize our
conclusions.
We demonstrate that such large quantities of absorbing material can
have a significant impact on the observed X--ray properties of cluster 
cooling flows.

\section{Calculations}
\label{sec:calc}

\subsection{Inhomogeneous Cooling Flow Models}
\label{sec:models}

As an input to the transfer calculations, we have selected a number of
representative models for inhomogeneous cooling flows, some of which were 
presented previously in Wise \& Sarazin (1993).
These models were calculated using the numerical technique described 
in White \& Sarazin (1987) and a subset have been used previously
by Sarazin \& Graney (1991) to estimate the optical coronal line
emission from cooling flows.
The optically thin X--ray properties for some of these models have 
been discussed previously in Wise \& Sarazin (1993, hereafter Paper I)
while the effects of opacity due to cooling gas were discussed
in Wise \& Sarazin (1999, hereafter Paper II).
We note that the opacity of gas which is in the process of {\it cooling}
below X--ray emitting temperatures (as opposed to the accumulated 
{\it cooled} material described in this paper) was shown previously 
in Paper II to be insufficient to account for the observed excess 
absorption. These cooling flow models have been discussed in detail 
in these papers and, in the interest of brevity, we will only briefly 
describe their relevant characteristics here.
For a more detailed discussion, the reader is referred to these sources.

The set of models discussed in Papers I and II span a range of
gas deposition including both homogeneous models as well as models
with varying levels of inhomogeneity.
Homogeneous models, by definition, exhibit a single density and
temperature at any given radius.
It is important to realize that homogeneous cooling flow models cannot 
reproduce the observed properties of the excess absorption.
In a homogeneous model, all of the gas cools below X--ray emitting 
temperatures just within the sonic radius, $r_s$, which is the point 
where the inflow velocity exceeds the sound speed in the gas
(see \S2.1 of Paper I and references therein).
Values for the sonic radii are typically quite small
($r_s \lesssim 1$ kpc);
therefore, any accumulated cooled material would be confined to the
very innermost regions of the cooling flow.
In contrast, the spectral results of White et al.\ (1991) indicate that 
the absorbing material must cover a significant fraction of the cooling 
flow region ($r \sim 100$ kpc), and the spatially resolved spectral 
measurements of Allen et al.\ (1993) and Irwin \& Sarazin (1995) confirm 
this extended absorption.
Consequently, in choosing models for these calculations, we have
only considered inhomogeneous models which feature cooling material 
(and by implication {\it cooled} material) distributed over large 
spatial scales.

The models used in this paper are listed in Table~\ref{tab:models}
and assume a spherically symmetric, steady--state inflow
with only gravitational and gas pressure forces included.
The outer boundary is taken to be the ``cooling radius'',
$r_c$, which corresponds to the point at which the instantaneous
isobaric cooling time, $t_c$, is equal to the age of the cluster.
In all cases, the cluster age was taken to be $t_a = 10^{10}$ years.
As described in Paper I, the models are completely determined by three
quantities: the total inflow rate $\dot M_c$; the gas temperature 
$T_c$; and the mass deposition profile, $\dot \rho(r)$, which describes 
how inhomogeneous gas cools below X--ray emitting temperatures as a 
function of radius in the cluster.
Both $\dot M_c$ and $T_c$ are evaluated at the cooling radius $r_c$
and values of $\dot M_c = 100$ or $300$ $M_{\sun}$  yr$^{-1}$  and 
$T_c = 8.0 \times 10^7$ K were used for all models.

To specify the mass deposition profile, $\dot \rho(r)$, two forms for 
the mass inflow rate were used.
For most of the models, the simple two--phase
cooling flow model of White \& Sarazin (1987) was employed.
In this prescription,
the mass deposition profile is parameterized as
\begin{equation}
{\dot\rho(r)} = q \frac{\rho(r)}{t_c (r)}
= \frac25
~q ~\frac{\rho(r)^3 \Lambda[T(r)]}{P(r)}
\label{eq_4:2}
\end{equation}
where $t_c$ is the instantaneous isobaric cooling time,
$\rho (r)$ is the gas density, $T(r)$ is the gas temperature,
and $P$ is the gas pressure.
The total emission and cooling rate per unit volume in the gas is
$\rho^2 \Lambda ( T )$, where $\Lambda$ is the total emissivity
coefficient.
The parameter $q$ is referred to as the ``gas loss efficiency
parameter.''
The values of $q$ for each of the models are given in column 5 
of Table~\ref{tab:models}.

In addition to models using the White \& Sarazin mass deposition 
formula (eq.~\ref{eq_4:2}), one model was included which used a
form for the mass deposition profile derived by various groups from 
deconvolutions of cluster X--ray surface brightness distributions.
Fabian et al.\ (1984), Stewart et al.\ (1984), and Thomas et al.\ (1987)
have argued on the basis of these observations that the inflow rate varies
as $\dot M (r) \propto r$.
In the C300\_8\_fb model, this assumption was adopted, which implies a
mass deposition profile of the form
\begin{equation}
\dot\rho(r) = \frac{ \dot M_c}{ 4 \pi r_c r^2 } .
\label{eqn:fabian}
\end{equation}
We note that this model is quite similar to a White \& Sarazin model
with $q \approx 3.4$.

The various physical parameters for each of the models is
given in Table~\ref{tab:models}.
The models use a naming convention of C$xxx$\_$y$\_$zz$, where
$xxx = \dot{M}_c$ (in units of M$_\odot$ yr$^{-1}$),
$y = T_c$ (in units of $10^7$ K), and $z = 10 \times q$.
For the ``Fabian et al.'' style model with $\dot{M} \propto r$, 
$z =$ ``fb.''
Three of the models were presented previously in
Sarazin \& Graney (1991) and Paper I, and the second column 
in Table~\ref{tab:models} gives their identifications in the 
nomenclature of those papers.
The remaining columns give the cooling rate $\dot{M}_c$ and temperature
$T_c$ at the cooling radius, the gas loss parameter, the cooling radius
$r_c$, and the sonic radius $r_s$.
A value of $r_s = 0$ indicates a fully subsonic solution.

\subsection{X--Ray Emission}
\label{sec:xemiss}

The calculation of the X--ray emission from the gas is discussed in
considerable detail in Papers I and II. 
We included bound-bound, bound-free, free-free, and two-photon emission
from ions of the elements H, He, C, N, O, Ne, Mg, Si, S, Ca, Fe, and Ni.
Elemental abundances for these species were set at one half of the
solar values as defined by Meyer (1979).
The emissivities were calculated under the usual ``coronal limit''
conditions, as described in Paper I.
Of course, the one major exception to these assumptions was the
inclusion of the opacity of the gas.
We assume that the cooled material is at temperatures below $10^6$ K,
and produces no appreciable X--ray emission.
At the column densities of interest, fluorescent line emission by
the cooled gas is not expected to be significant.
However, if the columns of cold material in the absorbing
clouds were larger, fluorescent emission of Fe K lines might be
detectable (Churazov et al. 1998).

\subsection{X--Ray Opacity}
\label{sec:xopacity}

In calculating the opacity in the cooling flow, we have included
contributions from the ambient hot gas, the {\it cooling} hot gas, and
the accumulated {\it cooled} material.
Of these three components, the first two (opacity due to ambient
and cooling material) were discussed in detail in Paper II and are
treated in the same manner here.
Processes contributing to the opacity of the hot gas include resonant line
absorption, photoelectric absorption, and electron scattering.
As discussed in Paper II, processes such as resonant scattering in
the X--ray lines or electron scattering merely
redistribute photons within the cluster and produce no net loss
in the spatially integrated spectrum.
We have also shown previously that the photoelectric opacity in an 
inhomogeneous cooling flow model due to ambient and cooling intracluster 
gas is small with $\tau \lesssim 0.01$ (see Paper II).
Thus, the opacity from these gas phases cannot account for the observed 
excess soft X--ray absorption in cooling flows.

In the present paper, we include a new source of opacity associated
with the accumulated cooled gas.
We assume that this material is relatively cold, so that its
opacity is primarily due to photoabsorption.
Values for the photoionization cross-sections were taken from
an updated version of the Raymond \& Smith (1972) emissivity code
provided by John Raymond.
The opacity of the cooled material will depend somewhat on
its ionization state.
We have assumed that the cooled material is cold and neutral.
Although the physical state of the cooled material has not been
established by observations,
White et al.\ (1991), Ferland, Fabian, \& Johnstone (1993),
and Daines, Fabian, \& Thomas (1994) have
argued that the absorbing material must be in the form of small, cold
clouds to have avoided detection at other wavelengths.
As mentioned above, the gas must be at a temperature $\lesssim 10^{6.4}$
for oxygen to retain K--shell electrons and thus present
a significant photoelectric cross-section (Shull \& Van Steenberg 1982).
In fact, the total opacity should remain fairly constant for temperatures
much below this value, since the cross-sections are fairly insensitive to 
changes in the ionization state once the ion has regained a full complement 
of K--shell electrons.

For a given model, we must also specify the amount and spatial distribution 
of the cooled material in order to calculate its contribution to the 
photoelectric opacity.
As discussed in \S\ref{sec:intro}, our approach is to construct
self-consistent models in which the absorbing material is assumed to
represent a fraction of all the gas which has cooled below X--ray 
emitting temperatures over the lifetime of the cooling flow.
Thus, we have taken the density of accumulated cold absorbing material 
as a function of radius in the cluster to be
\begin{equation}
\rho_{abs} (r) = \eta ~\dot \rho(r) ~t_a \, ,
\label{eq:rhocold}
\end{equation}
where $\dot \rho (r)$ is the rate at which material is cooling
below X--ray emitting temperatures and $t_a$ is the lifetime of
the cooling flow, taken here to be $10^{10}$ years.
The factor $\eta$ is a free parameter which allows us to vary
the fraction of the cooling material which is stored as cold 
absorbing gas. Since it is essentially a deposition efficiency, 
we assume that $\eta \le 1$.

Inherent in the adoption of equation~(\ref{eq:rhocold}) is the
assumption that the cold material accumulates where it cools.
Obviously, this assumption is a simplification, since in practice
a cloud of gas which is colder (and thus denser, assuming
pressure equilibrium) than the ambient gas will tend to fall in
towards the center of the cluster.
White \& Sarazin (1987) have calculated the distribution of the cooled
material assuming it moves on ballistic orbits.
Because such material spends a significant fraction of its orbital period
at the largest distances from the cluster center, the resulting distributions
are reasonably approximated by equation~(\ref{eq:rhocold}).
Alternatively, Nulsen (1988) and Fabian et al.\ (1991) have argued
that the cold clouds must comove with the background cooling flow.
Such co-motion also results in a distribution similar to
equation~(\ref{eq:rhocold}).
In any event, the likely departures from equation~(\ref{eq:rhocold}) 
are in the sense that the profile of the cold gas is more centrally 
condensed, which would tend to magnify the effects discussed in 
this paper.

\subsection{Numerical Solution of the Transfer Equation}
\label{sec:numeric}

We have solved the transfer of radiation through the cooling
flow numerically using the basic algorithm described by
Yorke (1980; 1986; 1988).
The transfer equation is cast in the cylindrical $(b,z)$ coordinate
system of Hummer \& Rybicki (1971).
Note, we use $b$ rather than $p$ to represent the impact parameter
which is fixed for a given ``ray'' through the
flow and corresponds to a given projected radius on the sky.
The coordinate $z$ measures the depth into the cluster along each ray.
With this choice of coordinates, the transfer equation can be written
\begin{equation}
\frac{d I_E(b,z)}{d z}
= \displaystyle{\epsilon_E - \kappa_E I_E} ,
\label{eq:ray}
\end{equation}
where $I_E$ is the intensity of radiation at photon energy $E$ at 
any point in a given direction along a ray.
The emissivity, $\epsilon_E$, contains contributions due to scattering
and therefore depends implicitly on integrals of the intensity $I_E$.

Having specified the variation of $\epsilon_E$ and $\kappa_E$ with
radius within the flow, we solve equation~(\ref{eq:ray}) using the
same numerical code employed in Paper II to calculate the effects
of opacity due to the cooling gas.
Spherical symmetry and a steady radiation field are assumed.
The code is non-relativistic and assumes that velocity $v$, temperature 
$T$, and photon energy $E$ satisfy
$| v | \ll c$, $ k T \ll m_e c^2$, and $E \ll m_e c^2$.
The effects of the flow velocity due to the cooling flow have been
included. This code is described in detail in Paper II and Wise (1992).
For relevant details of the numerical technique, the reader is
referred to these sources.

\section{Model Results}
\label{sec:results}

The models presented here are limited to the cooling flow region 
of the cluster ($ r \le r_c$).
In practice, the cooling flow region is immersed within and smoothly
connected to the general intracluster gas which extends out at least 
another order of magnitude in radius.
We have chosen not to include the ambient cluster component to our models
in order to focus on the properties of the cooling flow region and avoid
introducing additional parameters associated with the cluster
distribution.
In practice, this decision means that some of the results of our models will
differ from the equivalent properties directly observed in real clusters.
For example, the X--ray surface brightnesses in our models drop to zero 
at the cooling radius (e.g., see Figure~\ref{fig:spect_tot} below).
Thus, to compare these profiles with actual cluster surface brightness 
profiles, the emission due to the foreground and background cluster gas 
must be removed.

\subsection{Total X--Ray Luminosity}
\label{sec:xlum}

\begin{figure}
\vspace{6.50in}
\includegraphics{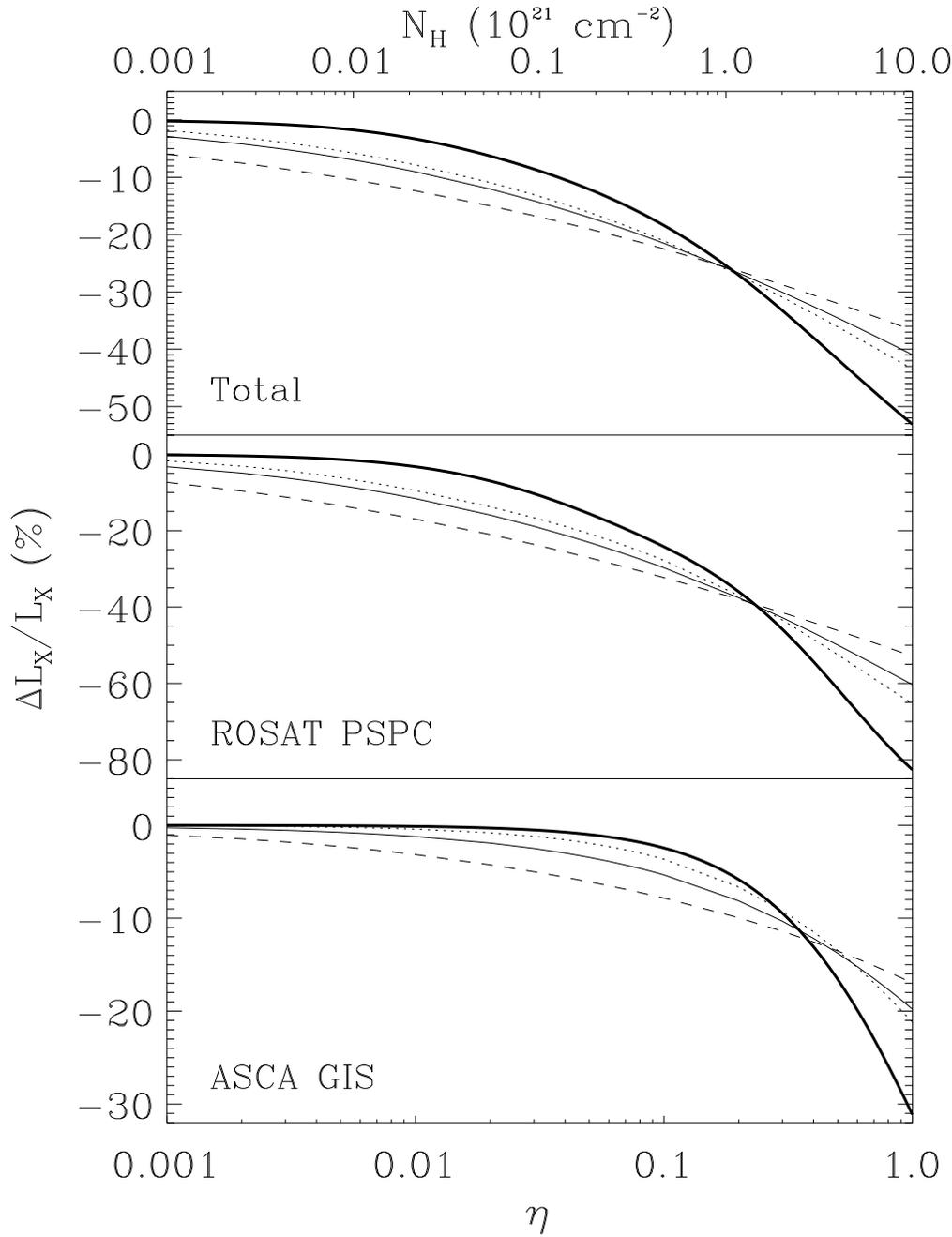}
\caption[Decrease in Total Flux for Absorbed Models]{
The percentage reduction in the X--ray luminosity as a function
of the efficiency parameter $\eta$.
The top panel corresponds to the total luminosity from 0.1--10 keV.
The middle and bottom panels show the reduction in the {\it ROSAT} PSPC and
{\it ASCA} GIS bandpasses, respectively.
In the three panels, curves are shown for several different
inhomogeneous models considered in the text.
The solid line corresponds to a model with a moderate $q$ value
of 1.0 while the dashed and dotted lines correspond to the 
extremes of the model grid.
The dashed curve shows the reduction for a model with $q=0.1$ while
the dotted line shows a model with $q=4.0$.
The heavy solid line in each panel indicates the percentage reduction
in the various luminosities for a simple foreground absorption model
with varying column densities $N_H$. The abscissa for the foreground
absorption curves is indicated along the top axis of the plot.}
\label{fig:flux_tot}
\end{figure}

Absorption due to the accumulated cold material will diminish the total 
amount of X--ray emission from the cooling flow.
The magnitude of this reduction in the total cooling flow luminosity
is shown in Figure~\ref{fig:flux_tot} as a function of the deposition 
fraction $\eta$.
Curves are given for three cooling flow models with
a small degree of inhomogeneity $q = 0.1$
(C300\_8\_01),
a moderate degree of inhomogeneity $q = 1.0$
(C300\_8\_10), and
a high degree of inhomogeneity $q = 4.0$
(C300\_8\_40).
Three different bandpasses are presented, corresponding to the total
X--ray flux (all photon energies $> 0.1$ keV), the flux observed with the
{\it ROSAT} PSPC (Position Sensitive Proportional Counter) instrument,
and the harder bandpass of the {\it ASCA} GIS (Gas Imaging Spectrometer).
Also shown for comparison are the effects of a column $N_H$ of
foreground absorption.

For values of the deposition fraction $\eta \sim 1$, the opacity
due to the accumulated cold material can produce a significant
reduction in the total X--ray luminosity on the order of 40\%, and
an even larger reduction in the flux seen in the soft {\it ROSAT} PSPC
bandpass (as large as 60\%).
The excess absorption columns seen by White et al.\ (1991) correspond to
values of $\eta \approx 0.1$ to 0.4 (see Figure~\ref{fig:ncol_eta} and
\S\ref{sec:spect_fit} below).
We note that these smaller values of $\eta$ still result in appreciable 
reductions of $\sim$30\% in the total flux and $\sim$45\% in the softer 
{\it ROSAT} band.

The differences between the three models in Figure~\ref{fig:flux_tot}
are relatively small.
At first sight, this lack of distinction is surprising, since a more 
homogeneous model ($q=0.1$, C300\_8\_01) deposits more absorbing
material in the center of the cluster where it covers less of the 
cooling flow region.
On the other hand, a more homogeneous model will also produce more of
its X--ray emission in this region as well.
In fact, equation~(\ref{eq:rhocold}) implies that there is always a
correlation between the distribution of the X--ray emissivity of the
cooling gas (proportional to $\dot \rho$) and the opacity of the
absorption due to cooled gas.
This correlation decreases the model-to-model differences in the reduction 
in the X--ray luminosity.

\subsection{Emergent X--Ray Spectra}
\label{sec:xspect}

\begin{figure}
\vspace{4.00in}
\includegraphics{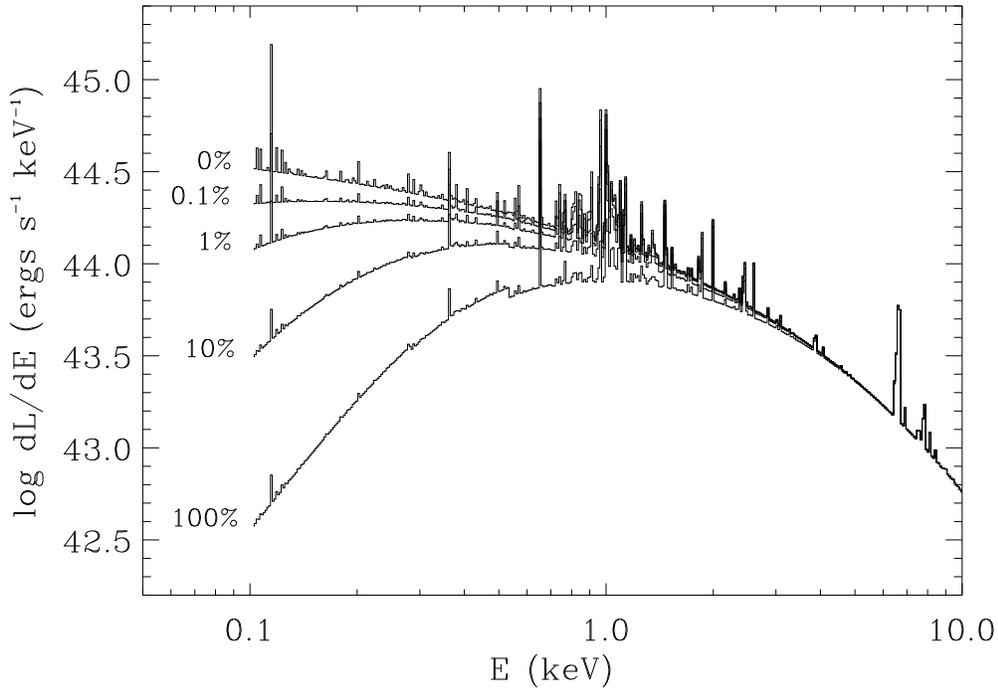}
\caption[Integrated Spectra for Absorbed Models]{
Total integrated X--ray spectra as a function of photon energy
for the $q=0.1$ (C300\_8\_01) model.
Curves are shown for differing values of the efficiency parameter,
$\eta$, described in the text.
A value of $\eta = 0$ corresponds to the the optically thin case.
Each curve is labeled with the corresponding value of $\eta$.}
\label{fig:spect_tot}
\end{figure}

The primary evidence for substantial amounts of cold, absorbing
material in the data of White et al.\ (1991) is a deficit of X--ray
emission for photon energies below about 2 keV over that expected
from purely galactic absorption.
To provide a basis for comparison with spectral observations
such as these, we have calculated the emergent X--ray spectra for 
our models. 
In Figure~\ref{fig:spect_tot}, the total integrated spectra
from 0.1--10 keV are shown for a fairly homogeneous model
($q=0.1$, C300\_8\_01) for varying values of $\eta$.
An absorption component due to foreground Galactic material
has not been included in this Figure.
As Figure~\ref{fig:spect_tot} demonstrates, the accumulated cooled
material does produce significant absorption below 2 keV even for
moderate values of $\eta$.

\begin{figure}
\vspace{4.00in}
\includegraphics{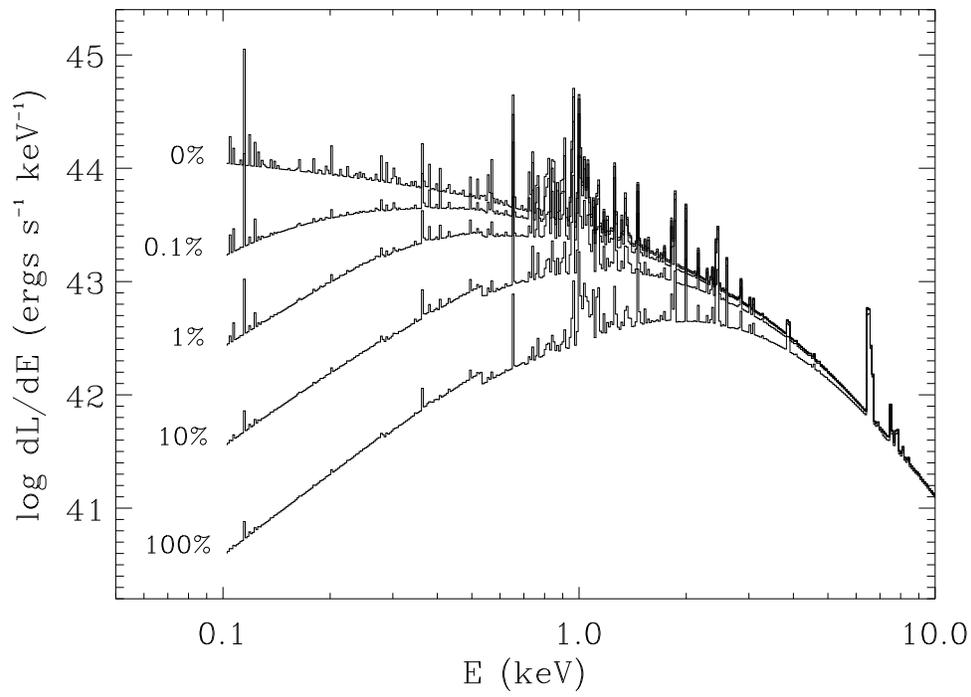}
\caption[Central Spectra for Absorbed Models]{
The central X--ray spectra as a function of photon energy
for the $q=0.1$ (C300\_8\_01) models.
This is the spectrum from a central circular aperture with a projected 
radius of 10 kpc.
The notation is the same as Figure~\protect\ref{fig:spect_tot}.}
\label{fig:spect_cent}
\end{figure}

For spatially resolved spectroscopy, the spectral signatures of excess
absorption can be even more apparent.
Integrated spectra average over the entire range of absorbing 
column densities in the cooling flow resulting in a lower mean 
level of absorption. 
However, a smaller aperture centered on the cluster will average
over higher columns resulting in a greater mean absorption. 
Figure~\ref{fig:spect_cent} shows several spectra for the same model 
extracted within a 10 kpc radius circular aperture centered on the core. 
The degree of absorption is correspondingly greater than in 
Figure~\ref{fig:spect_tot}.

In addition to the level of soft X--ray absorption, it is also
important to note that the accumulated cold material produces
a distinctive signature in the {\it shape} of the emergent spectrum.
Absorption due to foreground material produces an exponential fall off
in the observed flux toward lower energies since, at a given photon
energy $E$, $f(E)=f_o(E) \exp (-\tau_E)$ where $\tau_E$ is the optical
depth.  
However, for absorbing material which is intermixed with the emitting
gas, the dependence of the absorption on optical depth is weaker.
Roughly, we find that $f(E) \propto f_o(E)/\tau_E$.
Consequently for a given absorbing column, foreground absorption will
produce deeper absorption edges and a stronger decrease in the
spectrum toward lower photon energies than distributed absorption.

\begin{figure}
\vspace{4.00in}
\includegraphics{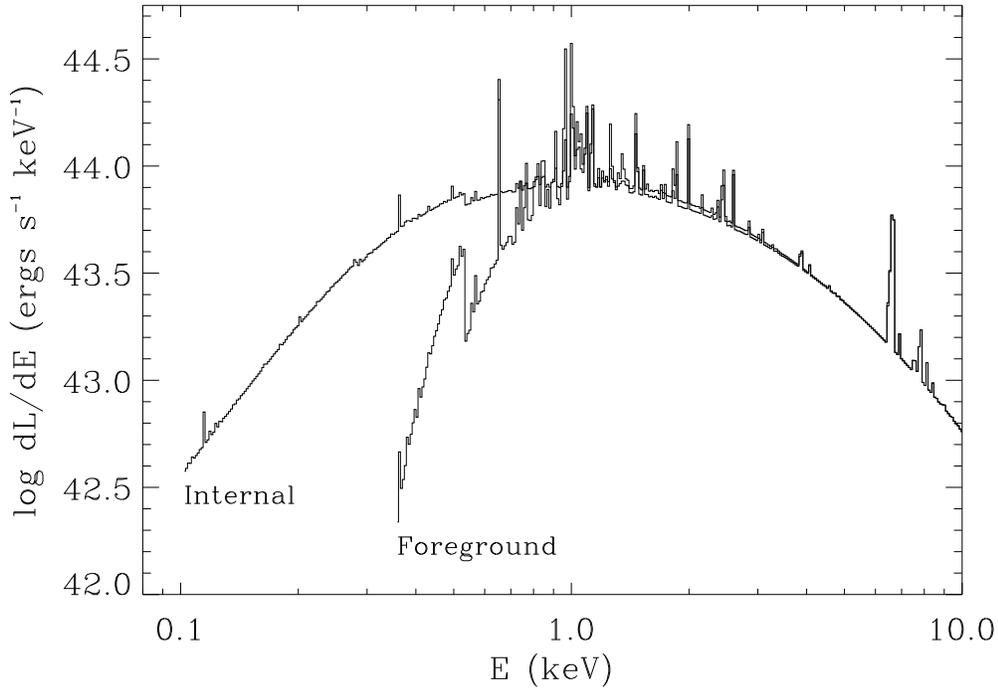}
\caption[Comparison of Interval vs.\ Foreground Absorption]{
The emergent X--ray spectrum from the $q=0.1$
(C300\_8\_01) model with
intrinsic absorption given by $\eta = 1$ is compared to
the same cooling flow model with no internal absorption, but
with a foreground absorption of $N_H  = 2.5 \times 10^{21}$ cm$^{-2}$.
Note that these spectra have similar shapes near 1 keV, but the foreground
absorption model falls off more rapidly at lower photon energies.}
\label{fig:spect_comp}
\end{figure}

This point is illustrated in Figure~\ref{fig:spect_comp}, which
compares the spectrum for the $q=0.1$ (C300\_8\_01) model with 
$\eta = 1.0$ (see Figure~\ref{fig:spect_tot}) with the spectrum 
of the same model with no internal absorption but foreground
absorption corresponding to $N_H = 2.5 \times 10^{21}$ cm$^{-2}$. 
This amount of foreground absorption was selected to give roughly the
same spectral shape as the internally absorbed model above 1.0 keV.
As noted, the model with foreground absorption decreases much more
steeply to lower photon energies, and has much deeper absorption edges
(e.g., the O K-edge at 0.532 keV).
In principle, this difference in the shape of the emergent spectrum
may provide a means to distinguish between foreground and intrinsic 
absorption.
For clusters with large foreground Galactic absorption, however,
making such a distinction may be problematic (see \S\ref{sec:spect_fit}).

\subsection{X--Ray Surface Brightness Profiles}
\label{sec:xsb}

\begin{figure}
\vspace{4.00in}
\includegraphics{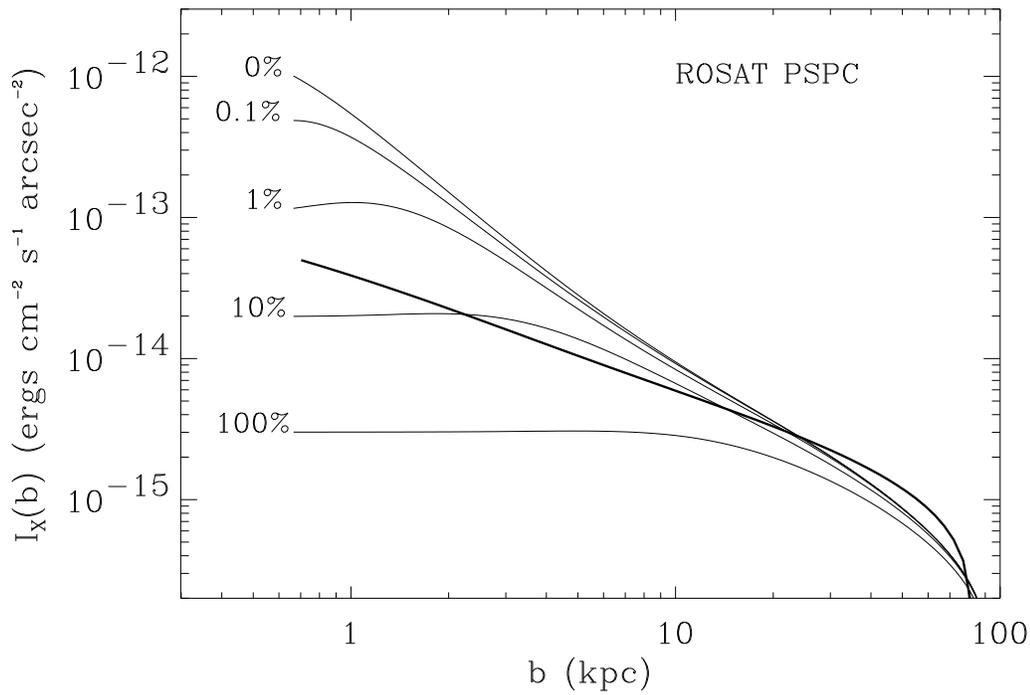}
\caption[{\it ROSAT} PSPC Surface Brightnesses for Absorbed Models]{
The emergent X--ray surface brightness profile in the {\it ROSAT} PSPC
bandpass as a function of projected radius for different values of the
efficiency parameter, $\eta$.
Curves are shown for the nearly homogeneous $q=0.1$ model
and $\eta = 0$ corresponds to the optically thin surface
brightness profile.
The heavy solid line represents the expected surface brightness
profile for a model with $\dot M(<r) \propto r$.
Each curve is labeled with the corresponding deposition efficiency
$\eta$ expressed as a percentage.}
\label{fig:sbtot}
\end{figure}

Centrally peaked X--ray surface brightness profiles represent one
of the primary pieces of evidence for the existence of cooling
flows in clusters of galaxies.
The X--ray emission due to the cooling gas is seen as an excess in the
central surface brightness profile relative to that of the outer parts
of the cluster (Jones \& Forman 1984).
In Figure~\ref{fig:sbtot}, the emergent X--ray surface brightness
profiles are shown for a fairly homogeneous model ($q = 0.1$, C300\_8\_01)
with values of $\eta$ ranging from 0 (no absorption) to 1.0 (100\%
of the accumulated cold material).
These profiles correspond to what would be observed in the {\it ROSAT} PSPC
bandpass. 
Clearly, the central peak in the cooling flow X--ray emission is 
strongly depressed by the presence of distributed absorption.

For comparison, Figure~\ref{fig:sbtot} also shows the X--ray surface 
brightness profile in the PSPC bandpass for the C300\_8\_fb model with
$\dot{M} \propto r$ and no internal absorption.
This model provides a reasonable fit to the observed X--ray surface
brightness profiles of many cooling flow clusters 
(e.g., Fabian et al.\ 1984).
As Figure~\ref{fig:sbtot} illustrates, even if only 10\% of the
cooling material collects in this cold form, substantial reductions in
$I_X(b)$ will occur. 
The rough resemblance of the intrinsically absorbed, essentially 
homogeneous model profile to that of the optically thin, 
$\dot{M} \propto r$ model highlights another important consequence of
distributed absorption. 
By suppressing the central peak of the surface brightness profile, 
a distributed absorber will make a given cluster cooling flow appear
more inhomogeneous than it actually is.

\begin{figure}
\vspace{4.00in}
\includegraphics{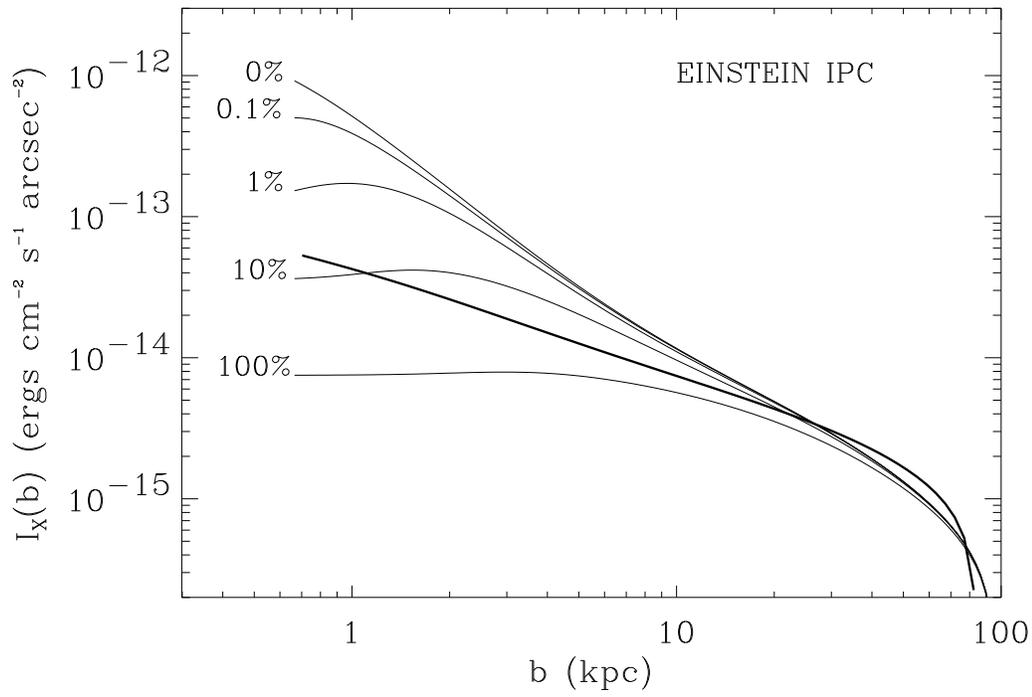}
\caption[{\it Einstein} IPC Surface Brightnesses for Absorbed Models]{
The emergent X--ray surface brightness profile as measured in the
harder {\it Einstein} IPC bandpass.
The notation is the same as Figure~\protect\ref{fig:sbtot}.}
\label{fig:sbein}
\end{figure}

{\it ROSAT} has provided high quality X--ray surface brightness profiles
for many clusters.
However, much of the initial work on the X--ray surface brightness
profiles of cluster cooling flows was based on observations with the
{\it Einstein} X--ray Observatory (e.g., Fabian et al.\ 1984).
The {\it Einstein} Observatory had a substantially harder X--ray
bandpass, extending to approximately 4 keV.
Since the effects of absorption (intrinsic or foreground) are lessened
at higher energies, one might expect {\it Einstein} surface brightness
profiles to be less sensitive to the effects of distributed absorption.
To test this intuition and provide standards for comparison to the
{\it Einstein} data, we have also calculated the X--ray surface brightness
for our models using the response of the {\it Einstein} IPC (Imaging
Proportional Counter). 
Figure~\ref{fig:sbein} confirms that internal absorption has a smaller
effect on $I_X$ as seen by the {\it Einstein} IPC than {\it ROSAT}.
Roughly speaking, one must increase $\eta$ by about a factor of two
to produce the same effect on {\it Einstein} observations as with {\it ROSAT}.
The qualitative effect on the surface brightness profile is the same.

\subsection{Resonant X--Ray Line Emission}
\label{sec:xline}

Strong X--ray line emission from cooling gas is one of the 
characteristic features of cluster cooling flows
(\markcite{claude88} Canizares, Markert, \& Donahue 1988).
Many of the X--ray lines produced by cooling plasma are non-resonance
transitions having small oscillator strengths, $f$, and correspondingly
small optical depths due to resonant absorption.
Such lines are affected by the internal absorption in the same way as 
the adjoining continuum discussed above. 
Consequently, the equivalent widths of optically thin lines are
not strongly affected by internal absorption.
However, many of the strongest lines arise from resonance transitions
and can be moderately optically thick in cooling flows (see Paper II). 
By itself, resonant line absorption (or scattering) merely
redistributes photons spatially and spectrally. 
When coupled with a significant source of continuum opacity, however,
resonant scattering can increase the effects of the absorption on
strong, resonance X--ray lines.
This increase occurs for lines with significant optical depth
because the effective pathlength through the absorbing medium is
increased. 
We consider some examples of these effects here.

\subsubsection{Line Equivalent Widths}
\label{sec:xline_ewid}

To illustrate some of the consequences of intrinsic absorption
for emission lines, we consider the observed properties of the Fe~XVII 
$2p^6{\phn}^1S-2p^53d{\phn}^1P^o$ resonance line which has a
wavelength of 15.01~\AA ~($E=0.826$ keV).
This line is produced by gas at temperatures in the range 
1--3 $\times 10^6$ K which is cooling rapidly out of the X--ray 
emitting regime.
It has a large oscillator strength ($f =2.35$) and an optical depth 
due to resonant absorption of $\tau \sim $0.3--2.0 depending on the
assumed cooling flow model (see Paper II).
Since this Fe~XVII line has a low photon energy, it is also
particularly susceptible to the effects of internal soft X--ray
absorption. 
In the simulations discussed here, we have assumed the line to be
thermally broadened.
This assumption maximizes the optical depth in the line.
The gas in cluster cores, however, may be quite turbulent 
($v_{turb} \sim 1000$ km s$^{-1}$) which would result in significantly
broader emission line profiles and lower line optical depths 
(see Paper II).

\begin{figure}
\vspace{4.00in}
\includegraphics{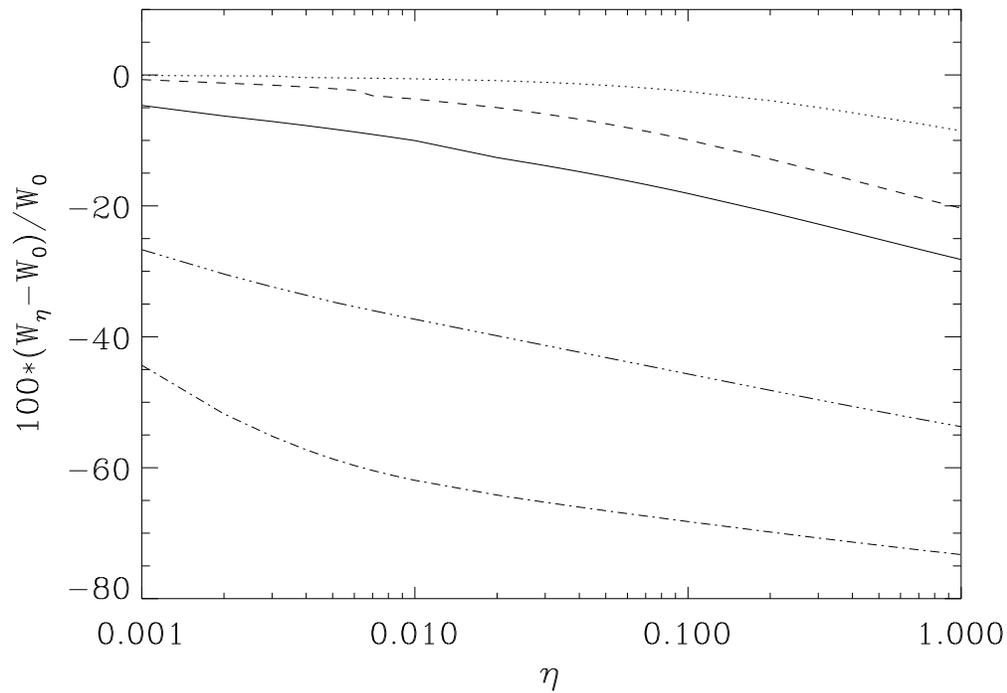}
\caption[Equivalent Width vs.\ $\eta for the Fe XVII Line]{
The percentage reduction in the total equivalent width, $W_{\eta}$, 
of the resonance Fe~XVII line at 0.826 keV as a function of the
deposition efficiency $\eta$ discussed in the text. 
Reductions are calculated relative to the optically thin equivalent 
width, $W_o$, for that model.
Curves are shown for varying values of $q$:
$q = 0.1$ (C300\_8\_01, dash--dot line), 
$q = 0.3$ (C300\_8\_03, dash--dot--dot--dot line),
$q = 1.0$ (C300\_8\_10, solid line), and
$q = 4.0$ (C300\_8\_40, dotted line).
The reduction for a model with $\dot{M} \propto r$ is also
included for comparison (C300\_8\_fb, dashed line).}
\label{fig:ew_fexvii}
\end{figure}

The reduction in the equivalent width, $W_{\eta}$, of the Fe~XVII line
as a function of $\eta$ is shown in Figure~\ref{fig:ew_fexvii} for
several models with varying degrees of inhomogeneity.
Models with $q$ values of 0.1, 0.3, 1, and 4 are shown ranging from
fairly homogeneous to strongly inhomogeneous, respectively.
The model with $\dot{M} \propto r$ is also included for comparison.
As Figure~\ref{fig:ew_fexvii} shows, resonant soft X--ray lines can be
strongly attenuated by internal absorption in cooling flows.
For a fixed value of $\eta$, the equivalent width of the line
decreases (relative to the optically thin value) as the deposition of
the cooling material becomes more condensed, i.e. as $q$ decreases.
Unlike the total luminosity of the flow, where the spatial coincidence
of absorbing and emitting material serve to reduce model--to--model 
variations, the increase in opacity due to resonant scattering
is more sensitive to the location of the absorbing material.
Thus, models with differing mass deposition profiles can produce very
different reductions in the line fluxes.
Note that the reduction in the equivalent width of the line in
Figure~\ref{fig:ew_fexvii} is in addition to the overall reduction in
the emission of continuum and optically thin lines at similar 
photon energies.
Clearly, the fluxes of these lines can be greatly reduced.

\subsubsection{Line Surface Brightness Profiles}
\label{sec:xline_sb}

\begin{figure}
\vspace{4.00in}
\includegraphics{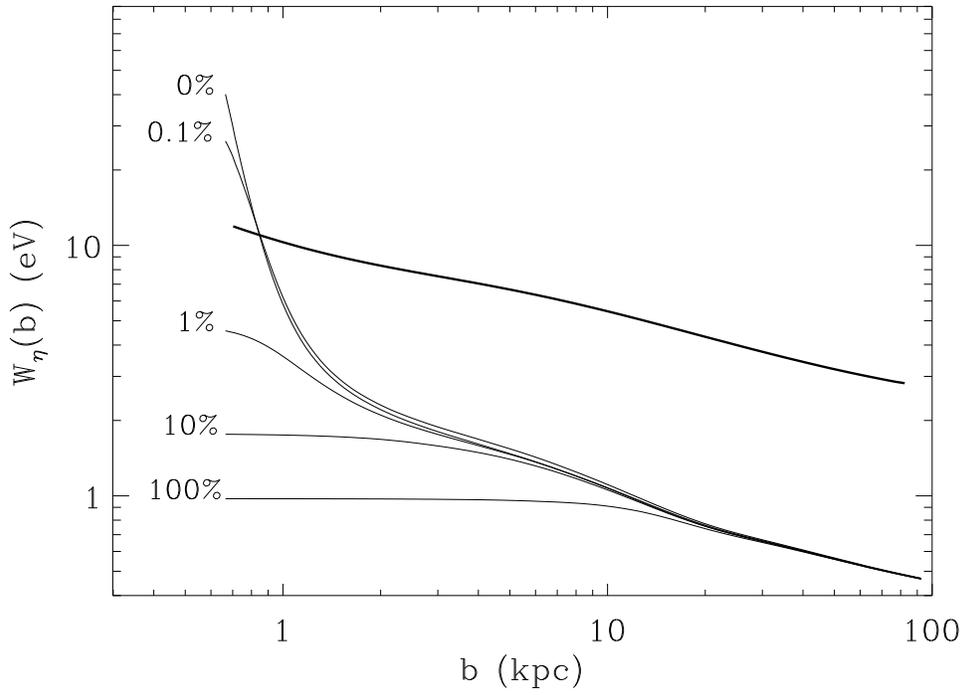}
\caption[Equivalent Width Profile for the Fe XVII Line]{
The equivalent width, $W_{\eta}$, of the resonance Fe~XVII line 
at 0.826 keV as a function of projected radius for the fairly
homogeneous $q = 0.1$ (C300\_8\_01) model assuming various values 
of $\eta$. Individual curves are marked with the value of $\eta$
assumed. The thick solid line shows the profile for the $\dot{M}
\propto r$ model with no internal absorption for comparison.}
\label{fig:ewpro_fexvii}
\end{figure}

Like their broadband counterparts, the presence of significant cold,
absorbing material will suppress the central peaks in the surface
brightness profiles of lower energy X--ray emission lines.
Profiles for resonance lines will be particularly affected due to
the interplay between absorption from the accumulated cold material
and the effects of resonant scattering.
Figure~\ref{fig:ewpro_fexvii} shows the emergent equivalent width 
of the resonance Fe~XVII line as a function of projected radius 
for the fairly homogeneous $q = 0.1$ (C300\_8\_01) model
assuming various values of $\eta$.
As the amount of internal absorption increases, the gradient in the
equivalent width flattens dramatically inside of $r \lesssim 10$ kpc.
Again, since the equivalent width compares the flux in the line to 
that of the adjoining continuum, the reduction in equivalent width 
is in addition to the overall reduction in the X--ray surface 
brightness shown in Figure~\ref{fig:sbtot}.

In the model without internal absorption ($\eta = 0$), the surface
brightness of the line is very strongly concentrated to the center
of the cluster. 
This concentration simply reflects the fact that, for relatively
homogeneous models, most of the gas cools inside the sonic radius
which occurs at rather small radii (see Paper I).
The Fe~XVII line is emitted by cooling gas at relatively low
temperatures, so the emissivity for this line is very large near the
center of the model.
Similar results are found for other resonance Fe L lines and for the
resonance K-lines of low-Z elements.
The K-lines of heavier elements (Si through Ni) are affected less
by the opacity of accumulated cooled material due to their higher
energy.

\subsubsection{Line Spectral Profiles}
\label{sec:xline_prof}

\begin{figure}
\vspace{4.00in}
\includegraphics{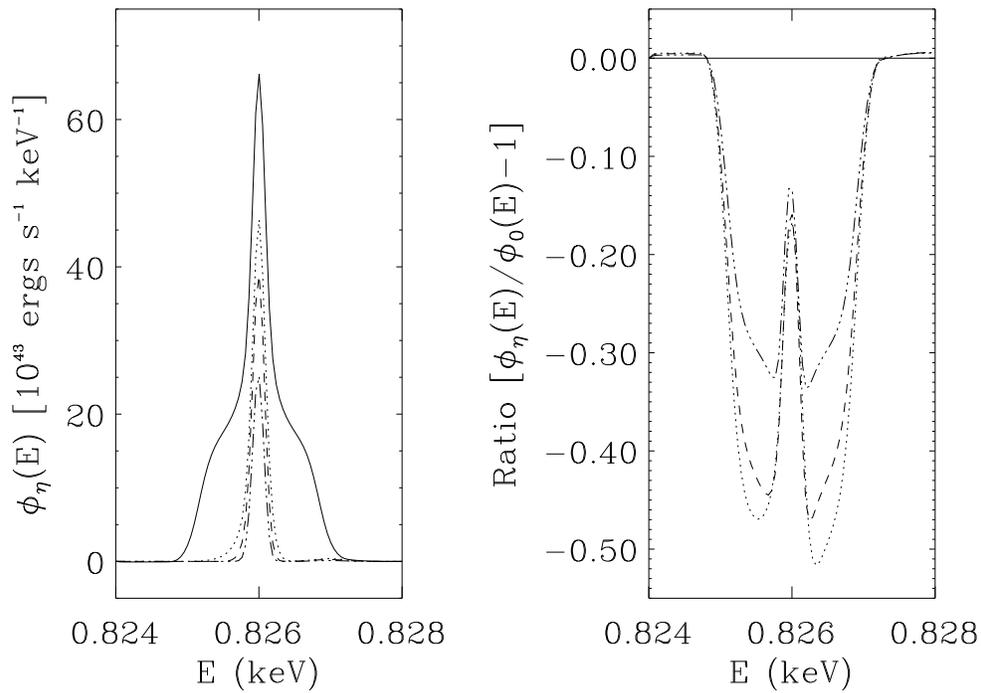}
\caption[Spectral Line Profiles for OVIII/Thermal/Central]{
The left panel shows the integrated spectral line profile, $\phi_{\eta}(E)$,
for the Fe~XVII line at 0.826 keV for different values of the efficiency
parameter $\eta$. These profiles correspond to a fairly homogeneous
$q=0.1$ (C300\_8\_01) model, and assume thermal line broadening.
The continuum flux has been subtracted from each profile.
The right panel shows the ratio of the emergent profiles to the 
model with no internal absorption.
Curves are shown for values of 
$\eta = 0.0$ (solid line),
$\eta = 0.01$ (dotted line),
$\eta = 0.1$ (dashed line), and
$\eta = 1.0$ (dash--dot--dot--dot line).}
\label{fig:lprofile}
\end{figure}

The combination of continuum opacity from accumulated cooled material
and resonance scattering will also alter the spectral line profiles of
low-energy resonance lines.
The left panel of Figure~\ref{fig:lprofile} shows the spectral
profile, $\phi_{\eta}(E)$, of the Fe~XVII line in the integrated
spectrum of a cooling flow model for different values of the
efficiency parameter $\eta$. These profiles were calculated for the
fairly homogeneous $q=0.1$ model (C300\_8\_01) and assume thermal line
broadening. The broad wings of the profile for the model without
internal absorption are due to gas moving radially inwards at a fairly
high velocity near the center of the model (see Paper I).

The right panel highlights the differences in the spectral profile
due to optical depth effects and shows the ratio of the profiles with
internal absorption to the profile with no accumulated cooled material.
Overall, the combined effects of resonance scattering and internal
absorption serve to reduce the strength of the line.
%
%
%
%
The broad wings of the line profile are produced by rapidly inflowing
gas near the center of the model, where the opacity is largest. 
Consequently, the wings are reduced more than the line center, where
there are contributions from larger radii.  
For lines produced by emitting material at larger radii or in models
with smaller inflow velocities, it is the line center which is most
strongly reduced.   
The spectral profiles depicted in Figure~\ref{fig:lprofile} also 
exhibit asymmetries relative to the line center.
These asymmetries arise because blue-shifted photons (due to material
flowing into the cluster center on the far side of the cluster) 
must traverse a greater absorbing column than red-shifted photons
(due to material flowing into the cluster center on the near side of
the cluster).

Similar results are found for other soft X--ray resonance lines.
Emission lines from higher ionization stages such as the Fe K$\alpha$
at 6.7 keV are less effected by photoelectric absorption, although
such lines can still be affected by resonance scattering (Paper II).
These effects are reduced if the lines are significantly turbulently
broadened, as this reduces the line scattering optical depth.
In any case, the detection of these effects will require very high
spectral resolution observations; as a result, it may be some time
before opacity effects on line shapes can actually be observed.

\subsection{Absorption Edges}
\label{sec:edge}

In addition to its broadband effects (see \S\ref{sec:xspect}),
accumulated cooled material in the cluster core will produce
absorption edges in the emergent X--ray spectra of cooling flows. 
At the columns of interest here, the strongest spectral feature
associated with the absorption is the O K-edge at 0.532 keV 
(e.g, Figure~\ref{fig:spect_comp}). 
If the intrinsic absorption is associated with the cooling flow,
the O K-edge will be redshifted to lower energies such that
$E_{edge} = 0.532/(1+z)$ keV.
Thus, two O K-edges will appear in the spectrum: an unredshifted edge
due to Galactic gas (and possibly oxygen in the detector and
spacecraft environment), and a redshifted edge. 
Detection of such a redshifted edge feature at the redshift of the
cluster would provide definitive evidence that the excess absorbing
material was associated with the cooling flow and not due to
anomalously high foreground absorption.
Arnaud \& Mushotzky (1998) have reported detection of such a
redshifted O K-edge in the Perseus cluster based on analysis 
of BBXRT spectra.

\begin{figure}
\vspace{4.00in}
\includegraphics{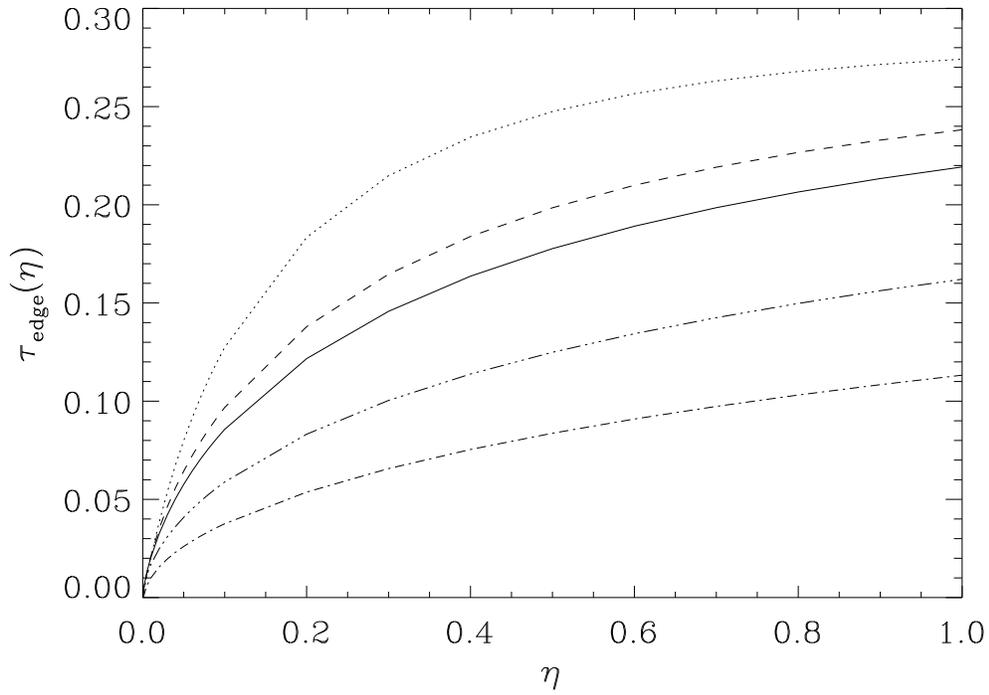}
\caption[Optical Depth in the O K edge vs.\ $\eta$]{
The optical depth in the O K edge at 0.532 keV in our models
is shown for different values of the deposition efficiency parameter
$\eta$. No foreground Galactic absorption has been included.
Curves are shown for models with
$q=0.1$ (dash--dot line),
$q=0.3$ (dash--dot--dot--dot line),
$q=1.0$ (solid line),
$q=4.0$ (dotted line),
and $\dot M \propto r$ (dashed line).}
\label{fig:edge}
\end{figure}

We can measure the observed depth of the O K--edge from our
self-absorbed cooling flow models directly.
In Figure~\ref{fig:edge}, the optical depth in the edge is plotted
for a range of deposition efficiencies, $\eta$, in the various models.  
The observed optical depth was determined assuming the decrement in the
observed flux at the edge energy was given by $\exp(-\tau_{edge})$.
Note that this assumption is equivalent to assuming a foreground
absorption model.
For $\eta = 1$, the measured depth in the edge ranges between 
$\tau \sim 0.1$--0.3 for all the models.
Given an absorption cross section at O K of 
$\sigma \sim 6.0 \times 10^{-22}$ cm$^{2}$, these optical depths imply
columns of $\Delta N_H \sim 2$--$5 \times 10^{20}$ cm$^{-2}$ assuming
the absorbing material is in the foreground of the cluster.
As we show below in \S\ref{sec:spect_fit}, this ``foreground''
assumption leads to significant underestimates of the true excess
absorbing column. 
No foreground Galactic column was assumed in Figure~\ref{fig:edge}.

\begin{figure}
\vspace{4.00in}
\includegraphics{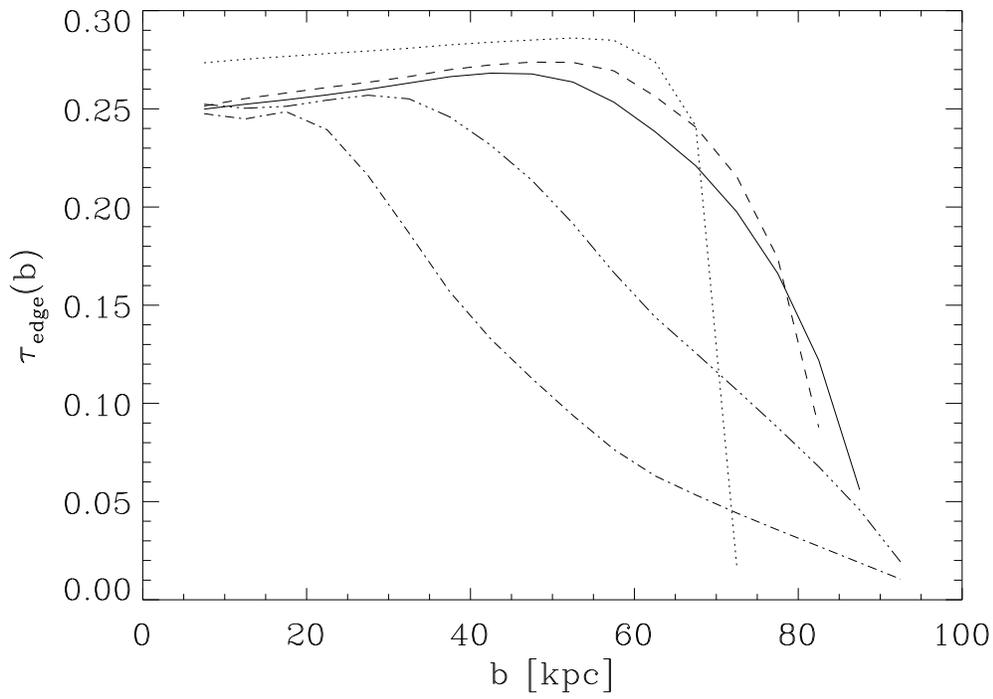}
\caption[Optical Depth in O K Edge vs.\ Projected Radius]{
The optical depth in the O K edge at 0.532 keV is shown
as a function of projected radius for our models.
The model spectra were accumulated in 5 kpc annular bins and a
value of $\eta=1$ was assumed.
The notation is the same as in Figure~\ref{fig:edge}.}
\label{fig:edge_spatial}
\end{figure}

The spatial variation of the O K-edge optical depth is shown in
Figure~\ref{fig:edge_spatial}.
For each model, the cooling region was divided into annular bins 5 kpc
wide and the optical depth in the edge was then determined for the
spectrum from each bin.
The different spatial profiles apparent in Figure~\ref{fig:edge_spatial}
simply trace the degree of inhomogeneity, and hence the degree to
which the absorbing material is centrally condensed, in the underlying
model.
For a given model, $\tau_{edge}$ increases steadily as one moves to 
smaller projected radii until the point where the cooling flow becomes
optically thick in the edge.
In the model with $\dot M (r) \propto r$, this point occurs at a
radius of $\sim 50$ kpc; while in the slightly inhomogeneous $q = 0.1$
model, $\tau_{edge}$ increases steadily into a radius of $\sim 10$ kpc. 
Since this ``saturation'' point corresponds to the radius at which the
actual optical depth in the edge is $\tau \sim 1$,
Figure~\ref{fig:edge_spatial} implies that the measured value of 
$\tau_{edge}$ can be used to estimate the true absorbing column 
at this radius.
From Figure~\ref{fig:edge_spatial}, the observed values for
$\tau_{edge}$ at these saturation radii range between $\sim 0.25$--0.30
implying underestimates of the true columns by factors of $\sim 3$--4.
This values are consistent with the results presented below based
on spectral fits to the integrated spectra assuming foreground
absorption models (see \S\ref{sec:spect_fit}).

In practice, it is difficult to separate an intrinsic, redshifted 
O K-edge from a Galactic edge with the {\it ASCA} SIS (Solid Imaging
Spectrometer) or {\it AXAF} ACIS ({\it AXAF} CCD Imaging Spectrometer)
because of their spectral resolution at these energies.
However, such observations may be possible with {\it AXAF} for some
higher redshift clusters ($z \ga 0.1$), where the intrinsic edge is
separated from the Galactic feature by $\ga 50$ eV.
Although the High Energy Transmission Grating (HETG) spectrometer on
{\it AXAF} has high spectral resolution, its throughput is low and
its resolution is degraded when observing extended sources.
Astro-E, on the other hand, should easily be capable of detecting
redshifted absorption edge features in cooling flow spectra.

\subsection{Scaling of Model Results}
\label{sec:scale}

Obviously, it is not possible to present results covering the full range
of all cluster cooling flow parameters in this study.
Thus, we have given a few characteristic examples to demonstrate the
effects of internal absorption.
However, it is generally possible to crudely assess the effects
of internal absorption for other parameter values by approximate
scaling of the models presented here.
The steady-state models we have presented are primarily characterized
by five parameters: the total cooling rate $\dot{M}_c$; the
temperature at the cooling radius $T_c$; the age of the cooling flow
$t_a$; the gas loss efficiency parameter $q$; and the fraction of
accumulated cooled gas which is present in an X-ray absorbing form $\eta$.
In addition, the models depend on the parameters characterizing
the gravitational field of the cluster and central galaxy, and the
abundances in the cooling gas.
However, if the temperature at the cooling radius is sufficiently high
and one is not concerned with the details of the innermost parts
of the flow near the sonic radius, the hydrodynamical solutions do not
depend strongly on the detailed form of the gravitational potential.
The atomic physics of the emission processes does depend in a complex
fashion on the abundances and the outer temperature, so we will assume
these are fixed in the scaling given below.

For a fixed temperature $T_c$ and fixed abundances, the density $\rho_c$
at the cooling radius $r_c$ depends inversely on age,
$\rho_c \propto t_a^{-1}$.
The total cooling rate in a model scales approximately as
$\dot{M}_c \propto \rho_c r_c^3 / t_a$.
Thus, the cooling radius in a model with fixed abundances and $T_c$
varies approximately as
\begin{equation}
r_c \propto \dot{M}_c^{1/3} \, t_a^{2/3} \, .
\label{eq:rc_scale}
\end{equation}
Equation~\ref{eq:rhocold} implies that the total mass of the cold
absorber in the model is
\begin{equation}
M_{abs}  = \eta ~\dot{M}_c ~t_a \, .
\label{eq:mcold}
\end{equation}
Consequently, the column density of the cold absorber $N_H (abs)$ and
the implied absorption optical depth $\tau_{abs}$ both scale as
\begin{equation}
\tau_{abs} \propto N_H (abs) \propto \frac{M_{abs}}{r_c^2}
\propto \eta \, \dot{M}_c^{1/3} \, t_a^{-1/3} \, .
\label{eq:abs_scale}
\end{equation}
Intervening absorption affects the observed properties in nonlinear
way; therefore, the value of $\tau_{abs}$ must be the same for two
models if they are to be scaled to one another.

On the other hand,  most of the emission properties of cooling flow
models nearly scale with the total cooling rate.
Thus, the luminosity in any spectral feature (and the flux in that
feature in the integrated spectrum) approximately scale as
\begin{equation}
L_\nu \propto \dot{M}_c \, .
\label{eq:lum_scale}
\end{equation}
The surface brightness of most features will scale approximately as
\begin{equation}
I_\nu \propto \frac{L_\nu}{r_c^2}
\propto \dot{M}_c^{1/3} \, t_a^{-4/3} \, .
\label{eq:surf_scale}
\end{equation}
Resonance emission lines present an exception to these scaling relations
(\S\ref{sec:xline}).
The properties of these lines depend on both the optical depth due
to absorption by cold material $\tau_{abs}$ and due to resonance
scattering $\tau_{rs}$. 
The latter scales approximately as
\begin{equation}
\tau_{rs} \propto \rho_c r_c \propto \dot{M}_c^{1/3} \, t_a^{-1/3} \, .
\label{eq:reson_scale}
\end{equation}
Requiring this quantity to be constant would generally make it
difficult to scale the models. In what follows, we will consider the
scaling of all properties of the models except those associated with
resonance emission lines which are optically thick to resonance scattering.

Let us assume one wishes to determine the properties of a model
characterized by $T_c$, $q$, $\dot{M}_c$, $t_a$, and $\eta$
which is not presented in this paper.
Further, assume that a model with the desired $T_c$ and $q$ is
presented, but with a different total cooling rate $\dot{M}_c^o$ 
and age $t_a^o$.
Then, from the grid of models presented in this paper, one should
select the model with a value of the deposition efficiency 
parameter, $\eta^o$, given by
\begin{equation}
\eta^o \approx \eta \,
\left( \frac{\dot{M}_c}{\dot{M}_c^o} \right)^{1/3} \,
\left( \frac{t_a}{t_a^o}  \right)^{-1/3} \, .
\label{eq:eta_scale}
\end{equation}
We note that this scaling can break down for models with large values
of $\eta$ and significantly different cooling rates or ages.
Strictly speaking, equation~\ref{eq:eta_scale} could produce values of
$\eta^o$ which are much greater than unity.  This much absorption
could not be produced by material cooling at the present rate.

Once the value of $\eta^o$ has been fixed, the luminosity and integrated
spectrum of the desired model can be scaled from the corresponding
fiducial model using
\begin{equation}
L_\nu \approx L_\nu^o  \, 
\left( \frac{\dot{M}_c}{\dot{M}_c^o} \right) \, ,
\label{eq:lum_scale2}
\end{equation}
while the surface brightnesses scale as
\begin{equation}
I_\nu \approx I_\nu^o  \, 
\left( \frac{\dot{M}_c}{\dot{M}_c^o} \right)^{1/3}
\left( \frac{t_a}{t_a^o}  \right)^{-4/3} \, .
\label{eq:surf_scale2}
\end{equation}
We have verified these relations for a number of cooling flow models,
and find that they work to within 10-20\% in essentially all cases.

\section{Effects of Intrinsic Absorption on Derived Quantities}
\label{sec:derive}

With the exception of Allen \& Fabian (1997), most previous analyses
of the X--ray spectra from cluster cooling flows have not considered
the effects of additional absorption due to accumulated cooled material.
Traditionally, any absorption in the cooling flow spectra has been
assumed to come from foreground material and in many cases has been
fixed at the Galactic value as measured by Stark et al. (1992).
Neglecting the effects of this material, if present, can lead to
serious errors in the derived physical properties of the gas in the
cluster core.
In the following discussion, we examine the effects of this intermixed
internal absorption on a number of physical quantities commonly
derived from X--ray observations.

\subsection{Total Cooling Rates}
\label{sec:mdottot}

Assuming the X--ray luminosity from a cooling flow is due to material
cooling isobarically from some temperature $T$, to first order
$L_X$ is simply proportional to the enthalpy of the cooling gas:
\begin{equation}
L_X \approx \frac52 \frac{kT}{\mu m_p} \dot M_c \, .
\label{eq:cool}
\end{equation}
For a measured total luminosity, $L_X$, this expression can be
inverted to obtain an estimate of the total cooling rate of material.
As we demonstrated in \S\ref{sec:xlum}, even moderate amounts of
accumulated cooled material can produce significant reductions in
the observed X--ray luminosity. 
Therefore, estimates of the total cooling rate $\dot M_{c}$ based
on the {\it observed} value of $L_X$  will tend to {\it underestimate}
the true value if the effects of internal absorption are neglected.
Based on the results in Figure~\ref{fig:flux_tot}, the derived values
of $\dot M_{c}$ might be as much as a factor of $\sim$2 too low.
As we show below (\S\ref{sec:spect_fit}), estimates of the X--ray
cooling rate based on fits to X--ray spectral observations which
assume only foreground absorption will tend to underestimate the 
true absorbing column and therefore $\dot M_c$ as well.

Individual X--ray emission lines can also be used to measure
the total cooling rate of material in cluster cores.
If the line is emitted primarily from material at temperatures below
the ambient cluster temperature $T_c$, then the luminosity in the line
is given by
\begin{equation}
L_{line} \approx \frac52 \dot{M}_c \frac{k}{\mu m_p}
\int_0^{T_c} \frac{\Lambda_{line} ( T )}{\Lambda (T)} \, d T ~~,
\label{eq:line_cool}
\end{equation}
where $\Lambda_{line} (T)$ is the emissivity coefficient of the line.
For lines emitted by gas at temperatures which are smaller than $T_c$,
the luminosity is nearly independent of $T_c$, the details of the
cooling, or even the abundances (which affect both $\Lambda_{line}(T)$
and $\Lambda (T)$ in a similar manner). 
This technique requires high spectral resolution observations
and has been applied to a number of clusters based on FPCS data 
from the {\it Einstein} X--ray Observatory (Canizares, Markert, \&
Donahue 1988). 
The High Energy Transmission Grating (HETG) onboard {\it AXAF}
should also provide data from several cluster cooling cores 
with sufficient spectral resolution to apply this technique.

For optically thin X--ray emission lines, the effects of intrinsic
absorption will reduce the observed luminosity in the line, $L_{line}$,
to the same degree as the nearby continuum.
Thus, lines at lower energies will suffer greater attenuation and
consequently underestimate the total cooling rate, $\dot M_c$, to a
greater degree relative to lines at higher energies.
For example, assuming a value of $\eta \sim 1$,
optically thin lines with energies near 0.6 keV could
suffer reductions as large as 60\% whereas the luminosity of a line 
at 2.0 keV would be reduced by only 20\%. 
These reductions are in addition to any reduction due to the
intervening Galactic absorbing column.
The actual observed reductions will of course depend on the true
underlying deposition profile of cooling material and the detector
response. 
If the line is a resonance line with an appreciable optical depth,
an additional reduction of about a factor of two is possible (see
\S\ref{sec:xline} and Figure~\ref{fig:ew_fexvii}).

As an aside, we note that differing amounts of intrinsic absorption
in cluster cooling cores may be partially responsible for the scatter
observed in correlations between $\dot M_{c}$ and other cooling flow
properties. 
Given the many potential sources of scatter in observed cluster
correlations, it seems unlikely that accumulated cold material
could be solely responsible.
However, as these models illustrate, the magnitude of these effects
is such that the presence of this material could represent a
significant contribution to the observed scatter in such correlations.

\subsection{Spatial Distribution of Cooling Material}
\label{sec:spatial}

Centrally peaked X--ray surface brightness profiles are one of the
characteristic features of cluster cooling flows.
However, it has been known for some time that the observed profiles
are less centrally peaked than would be expected if the gas in cluster
cooling flows was homogeneous and all of the gas flowed into the
inner regions ($r \la 1$ kpc) before cooling (Thomas et al.\ 1987;
White \& Sarazin 1987). 
The conventional explanation of this discrepancy is that the gas is
very inhomogeneous, and that denser lumps of gas cool more quickly.
Thus, a significant portion of the gas cools below X--ray emitting
temperatures before it flows in very far, and does not contribute
to the central X--ray emission.
One way of quantifying this model is to calculate the amount of hot
gas $\dot M (r)$ which is still flowing inward at each radius $r$.
By mass conservation, this quantity is equal to the amount of gas
which cools within this radius in steady-state.
Thomas et al.\ (1987) have argued that many clusters are reasonably
well represented by a mass deposition profile of the form 
$\dot M (r) \propto r$ out to $r \sim 200$ kpc.
One of our models (C300\_8\_fb) incorporates this spatial variation in
the cooling rate.

The implied form of the mass deposition profile $\dot M(r)$ will
depend on the observed X--ray surface surface brightness profile $I_X (b)$. 
However, as discussed in \S\ref{sec:xsb}, intrinsic absorption from
accumulated cooled material has the effect of reducing the central
surface brightness of cooling flows. 
Consequently, the gas in cooling flows may be less inhomogeneous
than a simple interpretation of the observed surface brightness
profiles might imply.
This effect is illustrated in Figure~\ref{fig:sbtot} which shows the
predicted {\it ROSAT} PSPC X--ray surface brightness profiles for a
slightly inhomogeneous $q = 0.1$ model with different values
of $\eta$ (expressed as percentages).
As mentioned previously, cooling flow models with $\dot M \propto r$ 
produce surface brightness profiles in reasonable agreement with
X--ray observations.
For comparison, the optically thin surface brightness profile for the
$\dot M \propto r$ model is indicated by the heavy solid line. 
Clearly, a value of $\eta \approx 10$\% is sufficient to bring the surface
brightness of this nearly homogeneous $q = 0.1$ model into rough
agreement with the optically thin $\dot M \propto r$ model. 
In \S\ref{sec:spect_fit} below, we show that this level of internal
absorption would correspond to an observed excess column of about $1
\times 10^{21}$ cm$^{-2}$ if the spectrum were fit assuming only
foreground absorption.  
This value is similar to the excess columns which have been derived
from X--ray spectra of cooling flows under this assumption 
(White et al.\ 1991).

\begin{figure}
\vspace{4.00in}
\includegraphics{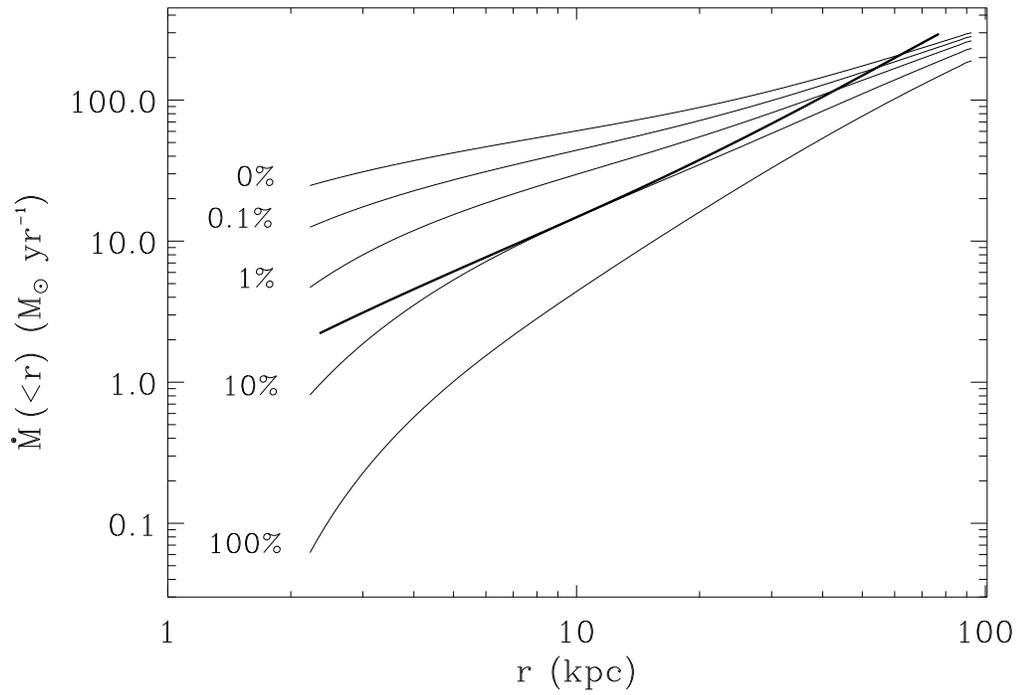}
\caption[Mass Deposition Profiles as a function of $\eta$]{
The deprojected mass deposition profiles, $\dot M(r)$, derived from the 
{\it ROSAT} PSPC X--ray surface brightness profiles calculated for
the $q=0.1$ model (C300\_8\_01) and shown in Figure~\ref{fig:sbtot}.
Each curve is labeled with the corresponding value of $\eta$ expressed
as a percentage.
For comparison, the heavy solid line shows the expected mass
deposition profile for a model with $\dot M (<r) \propto r$.}
\label{fig:mdot}
\end{figure}

As noted above, the variation of the inflow or cooling rate $\dot M (r)$
as a function of radius has been used to characterize the degree
of inhomogeneity of a cooling flow.
We have treated the predicted {\it ROSAT} X--ray surface brightness data
from our models as observed data, and deprojected the emission to determine
$\dot M (r)$.
Figure~\ref{fig:mdot} shows the deprojected mass deposition profiles
corresponding to the {\it ROSAT} X--ray surface brightness profiles
shown in Figure~\ref{fig:sbtot}.
For comparison, the heavy solid line shows the expected profile
for a model with $\dot M (r) \propto r$ (C300\_8\_fb).
As the degree of internal absorption is increased ($\eta \rightarrow 1$),
the derived $\dot M(r)$ profiles steepen which would be interpreted as
evidence that more of the gas is cooling below X--ray emitting
temperatures at larger radii.
Again, a value of $\eta \approx 0.1$ would cause the surface brightness
of this fairly homogeneous $q = 0.1$ model to have a mass deposition
profile which resembled $\dot M (r) \propto r$.

These qualitative results apply to all of the models.
We have illustrated the case for the fairly homogeneous $q = 0.1$ model
because the effects are particularly striking here.
However, we note that the specific values of $\eta$ required to turn a
more centrally condensed mass deposition profile (or equivalently,
surface brightness profile) into one which resembles the observed
$\dot M(r) \propto r$ will depend on the specifics of the input model
and the detector response. 
For example, due to its harder bandpass, if the above comparison 
were made with {\it Einstein} IPC data, an $\eta$ value a factor
of two larger would be required to produce the same effect.
Similarly, a much smaller spectrally determined excess column of order
$3 \times 10^{20}$ cm$^{-2}$ would convert a moderately inhomogeneous 
$q = 1$ model into a $\dot M (r) \propto r$ model.

As a final point, we note that the above effects do {\it not} apply
to completely homogeneous models. 
In such models, the gas does not cool to low temperatures until it 
is inside the sonic radius which typically occurs at very small
radii ($r \la 1$ kpc).
Therefore, any intrinsic absorption due to this accumulated cooled
material would also be concentrated within a very small region.
If present, this material would produce a large reduction in the
surface brightness profile but only at the very center of the cluster.
It would not affect the surface brightness at larger radii since no
absorbing material would have accumulated there.

\subsection{Absorbing Columns Estimated from Spectral Fits}
\label{sec:spect_fit}

The evidence for excess absorption in cluster cooling flows has come 
mainly from spectral fits such as those by White et al.\ (1991).
As we demonstrated in \S\ref{sec:xspect}, absorbing material intermixed 
with the emitting plasma produces a characteristic spectral shape in the 
emergent spectrum which is substantially different from the simple
exponential fall-off created by foreground absorption.
Consequently, fits to spectra from self-absorbed cooling flows
assuming only foreground absorption will yield erroneous values
for the derived columns.
Specifically, such fits will tend to {\it underestimate} the 
true absorbing column.
To quantify this effect, we have fit foreground absorption models 
to the emergent X--ray spectra of our models with internal absorption.
In the following analysis, we have used the effective area and 
spectral resolution of the {\it ROSAT} PSPC due to its good soft 
X--ray response.

\subsubsection{Fits to Integrated Spectra}
\label{sec:fit_total}

Most observations of excess absorption in cluster cooling flows have
been based on the integrated spectra for the entire cooling flow region. 
For example, the original study by White et al.\ (1991) utilized the
{\it Einstein} SSS detector with a field of view of 6 arcmin.
Most nearby cluster cooling flows are several arcmins in radius which
means that the SSS gives the integrated spectrum of the entire cooling
flow. 
Similarly, {\it ASCA} has a resolution of several arcmins and
typically yields integrated spectra for the entire cooling flow region.
The {\it ROSAT} PSPC has a better angular resolution of about 0.5
arcmin, but still only gives a few separate spectra within the cooling
flow regions of nearby clusters (e.g., Allen et al.\ 1993).
For comparison to such integrated spectra, we have determined the 
best-fit foreground column $\Delta N_H (spectral)$ using the
equivalent integrated spectra from our models.

To perform the spectral fits, we have used the following procedure.
For a given cooling flow model and choice of deposition efficiency,
$\eta$, we have calculated the integrated, emergent spectrum for the
entire cooling flow.
This spectrum was then convolved with the instrument response of the
{\it ROSAT} PSPC to obtain the number of ``observed'' counts per
energy bin, $F_{i}^{obs}$ for this model and choice of $\eta$.
The instrument response consists of the effective area, $A_{eff}(E)$,
and the redistribution function, $R(E,PHA)$, produced by the PSPC's
finite spectral resolution. 
As a fitting model, we used the optically thin ($\eta=0$) spectrum
for the selected cooling flow multiplied by varying amounts of
foreground absorption and convolved with the PSPC instrument
response.
For a given choice of $N_H$, this model yields the expected counts
per energy bin, $F_{i}^{fit}(N_H)$.
The value of $N_H$ which yielded the best fit was then obtained
by minimizing the $\chi^2$ statistic
\begin{equation}
\chi^2 = \sum_i
\frac{[ F_i^{obs} - F_i^{fit}(N_H) ]^2}{ F_i^{fit}(N_H) } \, ,
\label{eq:spec_fit}
\end{equation}
where $i$ is an index assigned to the spectral energy bins.
The $F_i^{fit}$ in the denominator of equation~\ref{eq:spec_fit} assures
the correct behavior if the errors in the observed spectra are dominated
by Poisson errors from photon noise.
Finally, because all clusters are observed through some amount of
Galactic material, both $F_i^{obs}$ and $F_i^{fit}$ were
also multiplied by varying amounts of Galactic foreground absorption
prior to fitting. 
Fits assuming Galactic columns of 0.1, 0.2, 1.0, and $2.0 \times 10^{21}$
cm$^{-2}$ were calculated and these columns were subtracted from the
best-fit foreground absorption values after the fit.

\begin{figure}
\vspace{4.00in}
\includegraphics{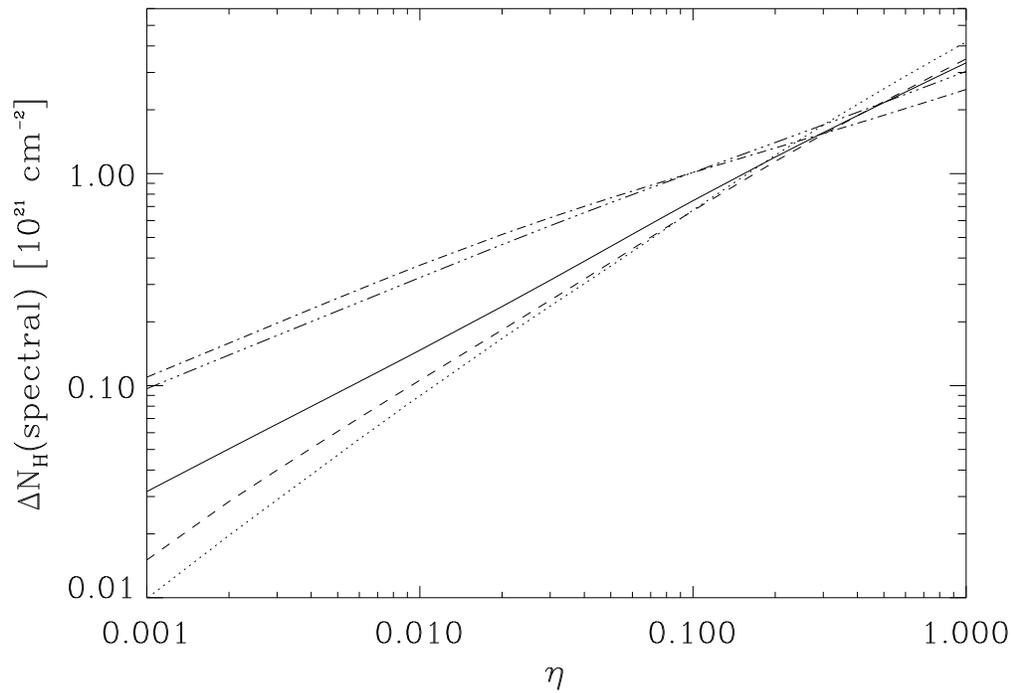}
\caption[Spectrally Determined Column Density vs.\ $\eta$]{
The excess absorbing column density for the various models
derived from spectral fitting assuming foreground absorption
is shown as a function of $\eta$.
The ordinate values represent the measured excess column after having
been folded through the {\it ROSAT} PSPC response. 
For all models, a foreground absorbing column of $N_H = 2.0 \times 10^{20}$
 cm$^{-2}$ was assumed and has been subtracted.
All of the models have $\dot{M}_c = 300 $~M$_\odot$ yr$^{-1}$ and
$T_c = 8 \times 10^7$ K.
Curves are shown for models with
$q=0.1$ (dash--dot line),
$q=0.3$ (dash--dot--dot--dot line),
$q=1.0$ (solid line),
$q=4.0$ (dotted line),
and $\dot M \propto r$ (dashed line).}
\label{fig:ncol_eta}
\end{figure}

Figure~\ref{fig:ncol_eta} gives the excess column densities determined 
using the above procedure as a function of the deposition efficiency $\eta$. 
Using this Figure, we can compare the properties of our models
with a given $\eta$ to the observed properties of clusters for which
the excess absorption $\Delta N_H (spectral)$ has been derived. 
Note that the values in Figure~\ref{fig:ncol_eta} are specific to 
$\dot{M}_c = 300 \, M_\odot$ yr$^{-1}$ and $t_a = 10^{10}$ yr.
The scaling to other values is given in equation~\ref{eq:abs_scale}.
The typical excess columns densities found by White et al.\ (1991)
assuming foreground absorption models are of order 
$\Delta N_H(spectral) \approx 1 - 2 \times 10^{21}$ cm$^{-2}$.  
These column values are in the regime where all of the model curves in
Figure~\ref{fig:ncol_eta} cross, and correspond to $\eta \approx 0.1$
to 0.4 for all of the models.

Observationally, values of $\Delta N_H(spectral)$ have been used to
derive limits on the total amount of cold, X--ray absorbing material
in cooling flows.
Consequently, it is useful to compare the spectrally derived columns
with those actually present.
The sense of the difference depends on whether one considers
spatially resolved spectra or the integrated spectrum of the entire
cooling flow region.
For spatially resolved spectra, the spectrally derived column
assuming only foreground absorption is always smaller than the
actual column along that line of sight.
This underestimate occurs because, for intermixed absorption, a
fraction of the absorbing material always lies behind a portion
of the emitting gas, and thus is not fully effective in producing
absorption.
However, if one considers the integrated spectrum of an extended
region where the absorbing column and emission vary with projected
position, the difference can have either sign.
A given amount of absorber is more effective if it is concentrated
where the emission is most intense, and such a concentration may not
be resolved in the integrated spectrum.

\begin{figure}
\vspace{4.00in}
\includegraphics{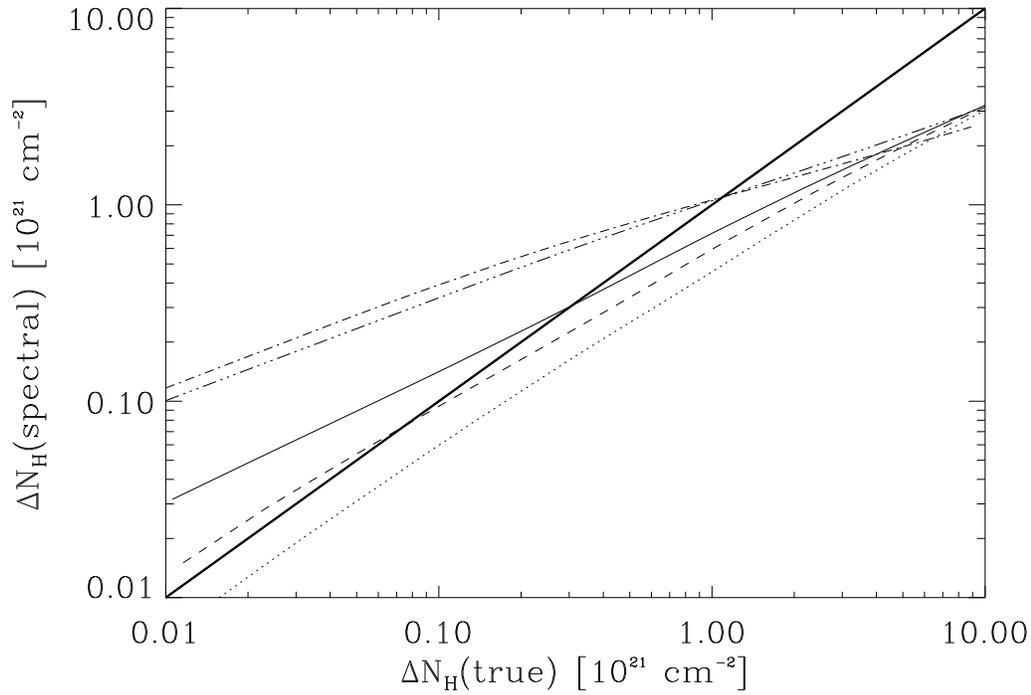}
\caption[Measured Column Density for Absorbed Models vs.\  Real]{
A comparison between the actual average internal absorption columns and the
values measured from fits to the spectra.
The actual average excess column is shown on the abscissa and is given by
$\Delta N_H (true) = {\dot M}_c t_a \eta / (1.47 m_p \pi r_c^2)$.
The ordinate values are the same spectrally derived columns shown in
Figure~\protect\ref{fig:ncol_eta}.
The models and notation are the same as in that Figure, except that
the heavy solid line indicates a one--to--one correspondence
between $\Delta N_H (true)$ and $\Delta N_H (spectral)$.}
\label{fig:ncol_comp}
\end{figure}

The spectrally-derived excess columns in Figure~\ref{fig:ncol_eta}
are determined from the integrated spectrum of the entire cooling
flow region.
Following equation~\ref{eq:abs_scale}, the average column density
of hydrogen in our models over the cooling flow region is given by
\begin{equation}
\Delta N_H (true) = \frac{M_{abs}}{1.47 m_p \pi r_c^2} =
\frac{\eta {\dot M}_c t_a}{1.47 m_p \pi r_c^2} \, .
\label{eq:dnh_real}
\end{equation}
Figure~\ref{fig:ncol_comp} compares the actual average column
densities determined using equation~\ref{eq:dnh_real} with those
derived from fits to the integrated spectrum of the cooling flow
region (Fig.~\ref{fig:ncol_eta}).
The heavy solid line indicates
$\Delta N_H (spectral) = \Delta N_H (true)$.
For low column in the inhomogeneous models, the spectrally-derived
columns overestimate the real column.
For large columns of intrinsic material,
$\Delta N_H (true) \ga 1 \times 10^{21}$ cm$^{-2}$, and for the
most inhomogeneous models, the spectrally derived column is smaller
than the real average column for all of the models by a factor $\la 4$.
For the more inhomogeneous models ($q \ga 2$) the measured columns
underestimate the true values down to columns of
$\Delta N_H (true) \sim 1$--$2 \times 10^{20}$ cm$^{-2}$.

These results can be compared to a similar analysis performed by 
Allen \& Fabian (1997) for a set of cooling flow clusters observed
with the {\it ROSAT} PSPC. 
Using {\sc XSPEC} simulations, these authors have simulated the errors
in the apparent excess column for two distributions of internal
absorption: a partial covering model and a ``multilayer'' model which
assumes a homogeneous distribution of absorbing throughout the X--ray
emitting region. 
These input models were then fit with a simple uniform foreground
absorption model similar to our procedure.
Of the two, this latter model is the closest approximation to the
detailed distributions of accumulated cooled material modeled here.
For values of $\Delta N_H (true)$ of 1 and $5 \times 10^{21}$ cm$^{-2}$
and a Galactic column of $N_H (gal)$ of $1 \times 10^{20}$ cm$^{-2}$,
these authors find that the fitted column densities underestimate
the true ``multilayer'' columns by factors $\sim 4$--5. 
These results are in good agreement with the factors of $\sim 3$--4
obtained from our models.
Allen \& Fabian (1997) also note that the degree of underestimation is
essentially independent of the Galactic column, a result we confirm.

For the more homogeneous models $q \la 2$, the spectrally-derived 
column can actually over-estimate the true average column for small
amounts of absorption.
This behavior is seen in Figure~\ref{fig:ncol_comp} for values 
of $\Delta N_H (true) \la 5 \times 10^{20}$ cm$^{-2}$.
In these models, both the emission and absorption are strongly
concentrated to the center of the cooling flow.
When the emitting and absorbing material are concentrated in this way,  
a smaller amount of absorber is more effective.
One way of understanding this effect is to note that in fairly
homogeneous models, the effective radii containing most of the
emission and absorption are both much smaller than the cooling
radius. 
Thus, if one averages the absorption over the projected area of the
entire cooling flow, as in equation~\ref{eq:dnh_real}, the resulting
value of the average column density $\Delta N_H (true)$ will be
much smaller than the true value in the region where the absorbing
material resides.

\subsubsection{Fits to Spatially Resolved Spectra}
\label{sec:fit_spatial}

With the exception of a few nearby clusters observed with {\it ROSAT},
the cooling cores of clusters have not been well resolved spatially by
instruments with good spectral resolution.
Consequently, studies of excess absorption have depended primarily
on analyses of integrated spectra.
However, {\it AXAF}'s excellent spatial resolution (FWHM $\sim 0.5$
arcsec) will provide spatially resolved spectra for many cooling flows.
With such future data in mind, we have calculated the values of 
$\Delta N_H (true)$ and $\Delta N_H (spectral)$ as a function of
projected radius in the cooling flow.
To determine $\Delta N_H (spectral)$, we divided the cooling region into 
annular bins 5 kpc wide. 
This choice gives roughly 20 bins over the cooling radius for each model.
For a choice of $\eta$, the emergent spectra within annular bins was 
calculated from the model surface brightness using
\begin{equation}
\frac{dL_E}{dE}(b_1:b_2) = 
\displaystyle{2 \pi \int_{b_1}^{b_2} db^\prime ~b^\prime ~I_E(b^\prime)} 
\label{eq:spect_ann}
\end{equation}
where $b_1$ and $b_2$ represent the inner and outer radii of the bins.
Similarly, emergent spectra were extracted from the corresponding 
optically thin model for the same set of annuli.
These spectra were then fit using the same procedure outlined above
to calculate the best-fit values of $\Delta N_H (spectral)$.

For comparison, the true variation of $\Delta N_H(b)$ was calculated
directly from the distribution of accumulated cooled material,
$\rho_{abs}(r)$ (see equation~\ref{eq:rhocold}), using
\begin{equation}
\Delta N_H(b) = \frac{1}{1.47 m_p}  \int \rho_{abs} ~dl = 
\frac{2}{1.47 m_p} \int_b^R dr ~\frac{
\displaystyle{r ~\rho_{abs}(r)}
}{
\displaystyle{\sqrt{r^2-b^2}}
}
\label{eq:dnh_spatial}
\end{equation}
where $l$ is the distance along the line of sight at a projected
radius $b$.
Given $\Delta N_H(b)$, the average column density within an annular 
bin was then taken to be
\begin{equation}
<\Delta N_H> (b_1:b_2) = 
\frac{
\displaystyle{2 \pi \int_{b_1}^{b_2} db^\prime ~b^\prime ~\Delta N_H(b^\prime)}
}
{
\displaystyle{ \pi (b_2^2 - b_1^2) }
} ~.
\label{eq:dnh_ave}
\end{equation}
We note that equation~\ref{eq:dnh_ave} reduces to 
equation~\ref{eq:dnh_real} in the limit that $b_1 \rightarrow 0$ and
$b_2 \rightarrow r_c$.
These expressions were used to derive the true average column densities
for the same annular grid used in determining $\Delta N_H (spectral)$.

\begin{figure}
\vspace{4.00in}
\includegraphics{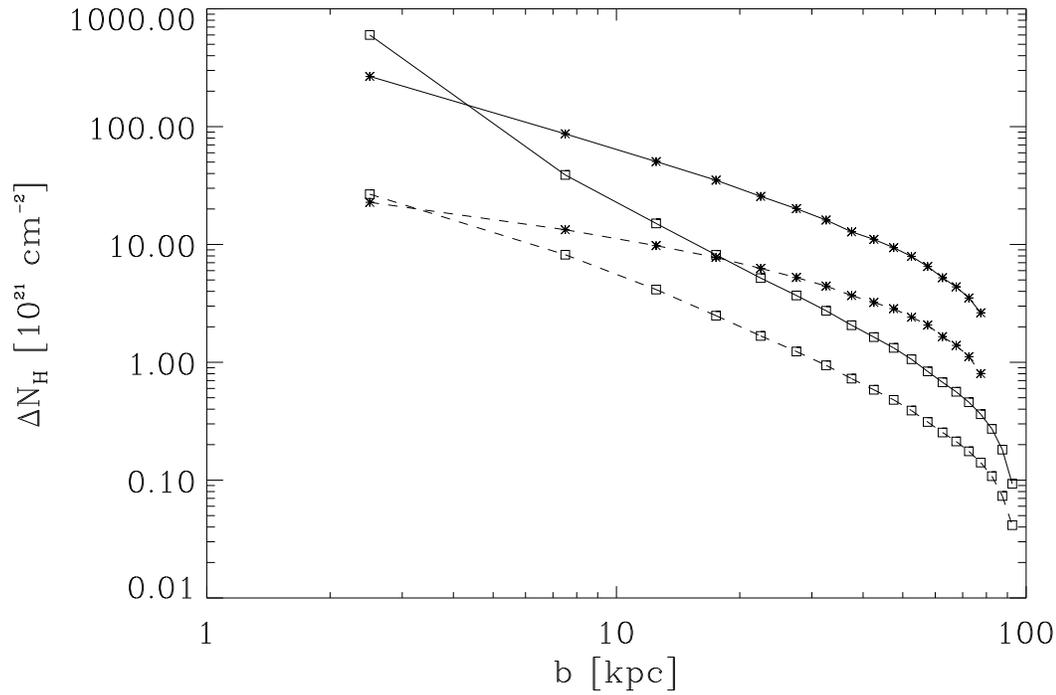}
\caption[Measured Column Density vs.\ Projected Radius]{
The variation in the true average internal absorption column and the 
fitted column as a function of projected radius for two cooling flow
models.
The slightly inhomogeneous $q=0.1$ model (C300\_8\_01) model is denoted
by square symbols while the model with $\dot M(r) \propto r$ (C300\_8\_fb)
is denoted by the stars.
The model spectra were accumulated in 5 kpc annular bins and fit with
a foreground absorption model as in Figures~\ref{fig:ncol_eta} and
\ref{fig:ncol_comp}. 
A value of $\eta=1$ was used in both cases.
For a given model, the solid line shows the true average column in the bin 
and the dashed line shows the resulting fitted values.
A foreground absorbing column of $N_H = 2.0 \times 10^{20}$ cm$^{-2}$ 
was assumed and has been subtracted.}
\label{fig:ncol_spatial}
\end{figure}

Figure~\ref{fig:ncol_spatial} shows the results of this spatial analysis
for two models: the slightly inhomogeneous $q=0.1$ model (C300\_8\_01)
and the model with $\dot M(r) \propto r$ (C300\_8\_fb).
As expected, the spectrally derived column density underestimates 
the true column at all projected radii.
The magnitude of this error increases as one moves towards the center
of the cluster.
For the model with $\dot M(r) \propto r$, the column determined from
spectral fits varies over a range of $\sim 3$--10 as one moves to
smaller radii.
In the $q=0.1$ model, more of the absorbing material is concentrated
toward the center of the flow producing underestimates of the true
absorbing column by factors of $\sim 2$--30.
As before, the Galactic column has been subtracted from the curves
in Figure~\ref{fig:ncol_spatial}.
{\it AXAF}'s spatial resolution should yield spatially resolved
spectra, and consequently measurements of $\Delta N_H$, on spatial
scales comparable to this example for clusters out to redshifts 
of $z \sim 0.1$.

\section{Conclusions}
\label{sec:conclude}

In this paper, we have calculated models for the X--ray properties of
cooling flow clusters including the effects of opacity due to
accumulated cooled material.
For consistency, we have assumed that the internal absorber has the same
distribution as the gas which is cooling below X--ray emitting
temperatures in the models.
We have characterized the amount of accumulated absorbing material as a
fraction $\eta$ of the amount of gas which would cool over the lifetime 
of the cluster ($t_a = 10^{10}$ yr) at the present rates of cooling. 
The emergent properties of the X--ray emission were then calculated
including the effects of radiative transfer through this accumulated
material.

The presence of accumulated cooled material will reduce the observed
X--ray fluxes and the implied cooling rates of cooling flow clusters
below the actual values. 
Depending on the efficiency with which cooling gas is converted
into absorbing material, the fluxes and cooling rates may be
underestimated by as much as a factor of two over the 0.1--10 keV
energy range.
For instruments with softer bandpasses, such as {\it ROSAT}, the
reduction can be higher.
Although modeling the effects of the intrinsic absorption with
foreground absorption models will reduce the magnitude of this
underestimate, these models tend to underestimate the total amount
of absorbing material and therefore do not completely compensate
for this effect.

Intrinsic absorption produces a characteristic spectral morphology
below $\sim 2$ keV which is not well fit by foreground absorption
models. 
Specifically, internal absorption does not increase in effectiveness
as an exponential function of the optical depth, as foreground
absorption does.
As a result, the integrated X--ray spectra for the models described here
do not decline as rapidly at softer energies and the absorption edges
are less deep than for models with foreground absorption.
For spatially resolved spectra toward the core of the cooling flow,
this effect is enhanced.
We find that these differences should be discernible in present {\it ASCA}
and future Astro-E and {\it AXAF} spectra.
On the other hand, it is difficult to distinguish spectra of models
with internal absorption from those with foreground absorption with a
partial covering factor.

In addition to its effect on the broadband X--ray spectrum,
accumulated cooled material will produce redshifted absorption edges
such as the O K--edge at an energy of $0.532/(1+z)$ keV.
These redshifted edges provide a direct means of determining whether
the absorbing material is intrinsic to the cooling flow and not due to
anomalously high foreground absorption. 
Resolving such redshifted O K--edge features from edges due to
foreground material and O in the CCD detectors of {\it ASCA} and 
{\it AXAF} is difficult for nearby clusters given the spectral
resolution of these instruments.
These features may be detectable in more distant clusters
with {\it AXAF} and {\it XMM} where the redshifted O K--edge 
is more cleanly separated or in nearby cooling flows with 
Astro-E which has greater spectral resolution.

The effects of internal absorption can have important implications
for the analysis of X--ray surface brightness profiles from cluster
cooling flows.
Because the absorbing material in our models is concentrated
toward the central regions of the cooling flow, it acts particularly to
diminish the central peak in the surface brightness and flatten the
observed profiles. 
Starting with {\it Einstein} observations, it has been
noted that cluster cooling flows are less centrally peaked than
would be expected if the cooling gas were homogeneous.
This fact has led to the canonical view that the gas in cooling flows
is rather inhomogeneous and that much of the gas cools at large radii.
However, our models indicate that opacity due to accumulated cooled
material can produce emergent surface brightness profiles much like
those observed even for nearly homogeneous gas distributions ($q=0.1$). 
Consequently, much more of the gas may be cooling below
X--ray emitting temperature in the central regions of cooling flows
($r \la 10$ kpc) than one would infer from observed X--ray surface
brightness profiles assuming the gas was optically thin. 
Another way of presenting the same result is to note that internal
absorption steepens the mass deposition profile, $\dot M (r)$, one
would derive from the observed surface brightness distribution,
making a nearly homogeneous model resemble one with
$\dot{M} (r) \propto r$.
We note that the reduction in the central surface brightness due to
internal absorption would also mean that central cooling times have been
overestimated, and that central gas densities and pressures have been
underestimated.

Low energy resonance X--ray emission lines are particularly sensitive
to the presence of internal absorption.
These lines can have appreciable resonance scattering optical depths in
cooling flows, and this scattering enhances the chances of absorption
by accumulated cooled material.
For lines such as the Fe~XVII line at 0.826 keV, the interplay between
continuum opacity and resonant scattering can lead to reductions of
20\%--80\% in the equivalent width of the line.
Like their broadband counterparts, estimates of the cooling rate
based on such lines will underestimate the true cooling rate. 
Line surface brightness profiles and spectral profiles are likewise
effected. 
With sufficient spectra resolution, one can resolve distortions in
the spectral profiles of these low energy resonance lines due to absorption.
However, detection of such signatures requires spectral resolution
well beyond that currently available.
Our models assume thermal line broadening. If significant turbulence is 
present in the cooling flow gas, the magnitude of these effects is reduced.
Emission lines from high ionization stages, such as the Fe K$\alpha$
line at 6.7 keV, are relatively unaffected by internal absorption though 
they may still have appreciable resonance scattering optical depths.

Previous studies of intrinsic absorption in cooling flows have
primarily determined the amount of absorbing material by fitting
foreground absorption models to the observed X--ray spectra.
As our models indicate, the spectral signatures of accumulated cooled
material are not well fit by such models.
For spatially resolved spectra, the spectrally derived columns always
underestimate the actual columns along the line of sight, typically by
a factor of $\sim 3$--20 depending on the specifics of the cooling
flow model.
If the amount of excess absorption is derived from integrated X--ray
spectra for the entire cooling flow (as has generally been the case to
date due to the limited spatial resolution of the spectrometers used),
the situation is more complicated.
For large columns or fairly inhomogeneous models, the absorption is
always underestimated this way.
However, for small columns in nearly homogeneous models, the degree
of excess absorption can be overestimated.
This effect occurs because both the emission and absorption in nearly
homogeneous models is strongly centrally concentrated, and a small
amount of absorbing material is more effective in this case.

\acknowledgements

C. L. Sarazin would like to thank A. Fabian for helpful conversations.
C. L. Sarazin was supported in part by NASA Astrophysical Theory Program
grant NAG 5-3057.
M. Wise's research at M.I.T. is supported in part by the {\it AXAF}
Science Center as part of Smithsonian Astrophysical Observatory
contract SVI--61010 under NASA Marshall Space Flight Center.

%
%
\clearpage
 
\begin{table}
\caption[Cluster Cooling Flow Models]{}
\label{tab:models}
 
\begin{center}
\begin{tabular}{lcccccc}
\multicolumn{6}{c}{Cluster Cooling Flow Models} \\
\hline \hline
Model & Previous & $\dot{M_c}$ & $T_c$ & Gas Loss & $r_c$ & $r_s $ \\
      & Model    & ({$M_{\odot}$} yr$^{-1}$) & (K)& & (kpc) & (pc) \\
\hline
C300\_8\_01&  &   300&$8.0 \times 10^7$&$q = 0.1$               &\phn95.6&      582\phd\phn\phn\\
C300\_8\_03&  &   300&$8.0 \times 10^7$&$q = 0.3$               &\phn93.5&      414\phd\phn\phn\\
C300\_8\_10&C3&   300&$8.0 \times 10^7$&$q = 1.0$               &\phn87.6&      115\phd\phn\phn\\
C300\_8\_40&C4&   300&$8.0 \times 10^7$&$q = 4.0$               &\phn72.3&\phn\phn0\phd\phn\phn\\
C300\_8\_fb&C5&   300&$8.0 \times 10^7$&$\dot{M} \propto r$&\phn83.7&   \phn11\phd\phn\phn\\
C100\_8\_10&  &   100&$8.0 \times 10^7$&$q = 1.0$               &\phn61.3&   \phn69\phd\phn\phn\\
\hline
\end{tabular}
\end{center}
\end{table}

%
%
%
%

%
%
\clearpage

%
%
\clearpage

%
%
\clearpage

%
%
\clearpage

%
%
\clearpage

%
%
\clearpage

%
%
\clearpage

%
%
\clearpage

%
%
\clearpage

%
%
\clearpage

%
%
\clearpage

%
%
\clearpage

%
%
\clearpage

%
%
\clearpage

%
%
\clearpage

\end{document}